\documentclass[onecolumn,draftcls, a4paper, 12pt]{IEEEtran}
\pdfoutput=1
\usepackage{latexsym}
\usepackage{graphicx}
\usepackage{comment}
\usepackage{amsfonts,amssymb,amsmath}
\usepackage{hyperref}
\usepackage[dvipsnames]{xcolor}
\def\BibTeX{{\rm B\kern-.05em{\sc i\kern-.025em b}\kern-.08em T\kern-.1667em\lower.7ex\hbox{E}\kern-.125emX}}
\newcommand{\EE}{\mathbb{E}} 

\newcommand{\ee}{{\rm e}}
\newcommand{\jj}{{\rm j}}  

\newcommand{\av}{{\bf a}}

\newcommand{\fv}{{\bf f}}

\newcommand{\pv}{{\bf p}}
\newcommand{\qv}{{\bf q}}
\newcommand{\rv}{{\bf r}}
\newcommand{\sv}{{\bf s}}
\newcommand{\tv}{{\bf t}}
\newcommand{\uv}{{\bf u}}
\newcommand{\wv}{{\bf w}}
\newcommand{\vv}{{\bf v}}
\newcommand{\xv}{{\bf x}}
\newcommand{\yv}{{\bf y}}
\newcommand{\zv}{{\bf z}}
\newcommand{\zerov}{{\bf 0}}
\newcommand{\onev}{{\bf 1}}

\newcommand{\Am}{{\bf A}}

\newcommand{\Hm}{{\bf H}}
\newcommand{\Id}{{\bf I}}
\newcommand{\Jm}{{\bf J}}
\newcommand{\Km}{{\bf K}}

\newcommand{\Mm}{{\bf M}}

\newcommand{\Qm}{{\bf Q}}

\newcommand{\Tm}{{\bf T}}

\newcommand{\Vm}{{\bf V}}

\newcommand{\Ym}{{\bf Y}}
\newcommand{\Zm}{{\bf Z}}

\newcommand{\Hc}{{\cal H}}

\newcommand{\Mc}{{\cal M}}

\newcommand{\Pc}{{\cal P}}

\newcommand{\Sc}{{\cal S}}

\newcommand{\Uc}{{\cal U}}

\newcommand{\alphav}{\boldsymbol{\alpha}}

\newcommand{\gammav}{\boldsymbol{\gamma}}
\newcommand{\deltav}{\boldsymbol{\delta}}

\newcommand{\etav}{\boldsymbol{\eta}}

\newcommand{\varphiv}{\boldsymbol{\varphi}}
\newcommand{\psiv}{\boldsymbol{\psi}}

\newcommand{\xiv}{\boldsymbol{\xi}}

\newcommand{\Gammam}{\boldsymbol{\Gamma}}

\newcommand{\Psim}{\boldsymbol{\Psi}}
\newcommand{\Thetam}{\boldsymbol{\Theta}}

\newcommand{\sinc}{{\hbox{sinc}}}
\newcommand{\diag}{{\hbox{diag}}}

\def\trace{\mathsf{Tr}}

\def\Herm{^{\mathsf{H}}}
\def\Tran{^{\mathsf{T}}}

\def\ben{\begin{enumerate}}
\def\beq{\begin{equation}}
\def\beqa{\begin{eqnarray}}
\def\bit{\begin{itemize}}
\def\een{\end{enumerate}}
\def\eeq{\end{equation}}
\def\eeqa{\end{eqnarray}}
\def\eit{\end{itemize}}

\def\non{\nonumber\\}

\newtheorem{remark}{Remark}
\newtheorem{proposition}{Proposition}
\begin{document}

\title{Optimization of IRS-aided Sub-THz Communications under practical design constraints}

\author{Alberto Tarable, Francesco Malandrino~\IEEEmembership{Senior~Member, IEEE}, Laura Dossi,
  Roberto Nebuloni, Giuseppe Virone,~\IEEEmembership{Senior-Member, IEEE}, Alessandro~Nordio,~\IEEEmembership{Member, IEEE}
\thanks{A. Tarable, F. Malandrino, L. Dossi, R. Nebuloni, G. Virone, A. Nordio are with
the National Research Council of Italy, Institute of Electronics, Information Engineering and Telecommunication  (CNR-IEIIT), 10129
Torino, Italy (e-mail: $<$name$>$.$<$surname$>$@ieiit.cnr.it).}}

\maketitle

\begin{abstract}
    We consider the optimization of a smart radio environment where
    meta-surfaces are employed to improve the performance of multiuser
    wireless networks working at sub-THz frequencies.  Motivated by
    the extreme sparsity of the THz channel we propose to model each
    meta-surface as an electronically steerable reflector, by using
    only two parameters, regardless of its size. This assumption,
    although suboptimal in a general multiuser setup, allows for a
    significant complexity reduction when optimizing the environment
    and, despite its simplicity, is able to provide high communication
    rates.  We derive a set of asymptotic results providing insight on
    the system behavior when both the number of antennas at the
    transmitter and the meta-surfaces area grow large.  For the
    optimization we propose an algorithm based on the Newton-Raphson
    method and a simpler, yet effective, heuristic approach based on a
    map associating meta-surfaces and users.  Through numerical
    results we provide insights on the system behavior and we assess
    the performance limits of the network in terms of supported users
    and spatial density of the meta-surfaces.
\end{abstract}

\begin{IEEEkeywords}
Intelligent Reflecting Surfaces, Multiuser channel, optimization, TeraHertz communications.
\end{IEEEkeywords}

\section{INTRODUCTION\label{sec:intro}} 
  \IEEEPARstart{T}{he} recent advent of the fifth generation (5G) of
  wireless mobile communications is revolutionizing the way we live
  and work, thanks to the massive increase of network capacity, to its
  ultra-low latency, and the possibility to connect hundreds of
  billions of devices. To achieve this goal, millimeter wave (mm-wave)
  communications combined with massive multiple-input multiple-output
  (mMIMO) techniques have been advocated for boosting the bandwidth
  and the spectral efficiency, respectively.  It is also expected that
  the future generation of mobile communications (6G) will exploit
  sub-THz/THz frequency bands (0.1–-10~THz)~\cite{Aluoini-mmw-thz,
    Akyildiz-THz} for indoor as well as outdoor applications involving
  both static and mobile users, and when very high data rates are
  required over short distances.  However, such frequencies suffer
  from high path loss, harsh propagation conditions, and
  blockages.
  
  An innovative solution to overcome the shortcoming of THz bands is
  defined by the umbrella term of smart radio environment
  (SRE)~\cite{DiRenzo}.  SRE is a dynamically configured environment,
  where the interaction between radio waves and objects can be
  controlled in a programmable way. By considering the propagation
  characteristics of the environment as an exploitable resource rather
  than a source of signal degradation, SRE can potentially
  revolutionize the classical paradigm of wireless networking. One of
  the most promising and realistic implementations of SRE makes use of
  software-defined intelligent reflecting surfaces (IRSs)~\cite{Wu},
  which could be integrated within the walls of a room or of a
  building.  An IRS is a two-dimensional {\em meta-surface}, composed
  of a large array of passive or active scattering elements, called
  {\em meta-atoms}~\cite{Liaskos, zhang2021active}, with a
  specifically designed physical structure and radiation
  pattern. Meta-surfaces can control, in a software-defined manner,
  the phase shifts applied by individual meta-atoms to the incident
  waves.  By smartly adjusting such phase shifts, the reflected
  signals can be either coherently combined at the intended receiver
  to increase the received power, or destructively combined at
  non-intended receivers to mitigate interference, thus realizing an
  energy-efficient beamforming, allowing high performance with lower
  transmission power.  When the line-of-sight (LoS) link between the
  transmitter and the receiver is absent or severely degraded,
  connectivity can still be granted by pointing narrow radio beams
  towards an IRS in LoS and configure it to point to the receiver, as
  exemplified in Fig.~\ref{fig:map}.  An overview of principles and
  challenges of IRSs for wireless communications can be found
  in~\cite{ElMossallamy}, whereas~\cite{survey} provides an extensive
  review ranging from their physical characterization to the
  discussion of design methodologies and existing prototypes and their
  applications to wireless communications. Also,~\cite{tutorial_2021}
  elaborates on aspects such as passive reflection optimization,
  channel estimation, and deployment from various communication
  perspectives.  It is expected that IRSs will become key actors in
  future wireless networks, by synergically interoperating with other
  network control strategies and signal processing
  techniques~\cite{Full-duplex, multicell,cooperative-transmission}.
  By realizing the SRE paradigm, IRSs can make THz communication
  technologies viable for application in a large class of indoor and
  outdoor scenarios, as envisioned by current 5G and future 6G
  standards~\cite{Liaskos,Liaskos-3,Alsharif2020}, Moreover, IRSs
  allow to move part of the intelligence of the system from the
  transceivers to the environment, while improving the
  spectrum-sharing capacity in multiuser communications, as shown
  in~\cite{Tan}.  IRSs can be applied to a wide range of propagation
  scenarios, ranging from GHz to THz frequencies~\cite{pan2020sum}. In
  the GHz bands, channels are characterized by rich scattering, so
  that the radio links from/to each IRS element are assumed to undergo
  independent fading.  In such a situation, the phase shifts applied
  by the IRS elements need to be jointly optimized according to some
  figure of merit, such as network throughput. Typical solutions from
  the literature resort to, e.g., alternating optimization, as done
  in~\cite{Wu-Zhang, twotimescale}. Such approach, however, requires
  the estimation of large channel matrices with independent entries,
  entailing communication overhead. Moreover, its complexity increases
  with the number of IRS elements~\cite{Full-duplex,multicell,
    HongbinLi-2,HongbinLi-1,Wu-Zhang}.  The picture changes
  dramatically for systems operating in the sub-THz/THz frequency
  bands. Indeed, although the wireless channel at such frequencies is
  not yet completely characterized, it is known that it becomes
  sparser and its main components are the LoS and some non-LoS (NLoS)
  reflected rays, while scattering and diffraction provide little
  contribution to the received signal
  power~\cite{Akyildiz2018,tarboush2021teramimo,Xing}.  Moreover, if
  beamforming is employed and the signal power is concentrated in a
  specific direction, the effect of multipath is further reduced.  In
  such a scenario IRS can be more easily optimized and they can be
  configured to macroscopically act as programmable mirrors.

\subsection{Novelties and contributions}
In this work, we pose the following questions: in a sub-THz
communication system, how efficient a simple optimization of IRS phase
shifts could be? Or, more precisely, what performance is achievable
when the optimization algorithm treats the whole IRS as a single mirror-like entity,
as opposed to a mere collection of meta-atoms?
We anticipate the surprising answer: in some conditions, simplicity and
optimality are not incompatible.

To answer these questions we consider the downlink of a wireless network
where a base station (BS) communicates with a set of $K$ randomly
placed user equipments (UEs) through $N$ IRSs. There is no direct path
from the BS to the UEs, however a NLoS path exists between them, due
to the reflection from a large wall.  Both the BS and the UE are
equipped with arrays of antennas so that they are able to apply
beamforming techniques. While the BS antenna array can perform digital
beamforming, we consider a practical scenario where, due to limited UE
cost, the UEs antenna array can only allow for analog beamforming. The
scenario is depicted in Fig.~\ref{fig:map}.

We adopt a sparse channel model, typical of sub-THz/THz
frequencies where multipath is due to few reflectors
(e.g., large static objects, such as walls) and the path loss is
characterized by both large-scale fading and molecular
absorption. 

We consider the IRS model in~\cite{Larsson_2020}, and since used by
several authors~\cite{Wymeersch, noiICC2020,scattermimo}
which describes an IRS as an array of subwavelength-sized diffuse
scatterers which phase-align their reflected signals in a specified direction
in order to achieve ``anomalous'' reflection properties.
The phase shifts applied by the meta-atoms are related
to each other through a linear equation. Finally, each IRS is characterized by two design parameters,
regardless of the number of meta-atoms, namely:
\begin{itemize}
\item a constant {\em phase gradient} among the IRS elements, which,
  according to the generalized Snell's law~\cite{Larsson_2020},
  determines the {\em steering angle} applied by the IRS to the impinging
  signal;
  \item a {\em phase shift}, which adjusts the phase of the signal
    reflected by each IRS in order to constructively/destructively
    interfere with other desired/undesired signals at the UE end.
\end{itemize}
Such IRS model is known to be optimal for a single-user system in pure
LoS IRS-UE channels although it is suboptimal for
a multi-user systems characterized by multipath channels.
Nevertheless, it has several advantages: first of all,
each IRS can be characterized by only two parameters, hence the
complexity of the optimization process is significantly
reduced. Secondly, the estimation and exchange of a large number of
channel coefficients, one for each IRS element, is not required, as
the system only needs the knowledge of the positions of the UEs to
serve. Also, it allows for the design of efficient SRE
optimization algorithms which, in a wide range of relevant conditions, are equivalent to a
smart one-to-one association between UEs and IRSs, as shown later.
In fact, for highly directive beams, as is the case in  sub-THz and THz applications, when IRSs are optimally configured 
to serve a certain UE, due to the asymptotic orthogonality
among channel vectors associated with different users, IRSs
will become interference-free to other users~\cite{HongbinLi-1}. 
Finally, the optimality of the model in~\cite{Larsson_2020} also holds 
for several multiuser setups, as shown later.

Our main contributions can be summarized as follows:
\begin{itemize}
\item we analyze the performance of the downlink in a practical
  IRS-aided multi-user network operating in the sub-THz frequency
  band, where the direct BS-US LoS path is blocked and multipath components
  are due to reflection on few large static objects;
\item we formulate an SRE optimization problem, considering a case in
  which the BS performs zero-forcing (ZF) precoding, the performance
  metric is the received signal-to-noise ratio (SNR) at the UEs,
  and the optimization variables are the phase gradients and the phase shifts
  for all IRSs, as well as the steering directions of the UEs antenna arrays;
\item we propose a simple and efficient heuristic solution, based on
  the Hungarian algorithm, which associates a IRS to each user; we
  then analytically show that this approach becomes essentially
  optimal in the asymptotic regime where the IRSs areas are large and
  the BS array has a large number of antennas;
\item through numerical analysis, we highlight the role of the network
  design parameters and their effects on the system performance; in
  particular (i) we assess the contribution to the received SNR due to
  NLoS components, (ii) we verify the optimality of the heuristic
  approach in realistic conditions, (iii) we assess the
        performance loss incurred when discrete phase-shifters are
        employed and (iv) we provide design rules for the sizing of
      the system, in order to be able to fully exploit the degrees of
      freedom that are intrinsically available in a reference
      scenario;
\end{itemize}
Beyond taking into account several aspects that are most often
overlooked in the literature (reflection by large static objects,
multiple antennas at the UEs), the novelty in the paper resides in the
proposal of an efficient, {\em practical} way of performing SRE
optimization, compatible with the low delay required by near-real-time
scenarios, and in the analysis of its performance, whose understanding
paves the way for the implementation of a veritable IRS-assisted
quasi-orthogonal space-division multiple access.

The reminder of the paper is organized as follows.  In
Section~\ref{sec:model}, we introduce the communication model and
characterize the IRSs. In Section~\ref{sec:asymptotic}, we provide an
asymptotic expression of the channel
matrix. Section~\ref{sec:rotation} derives the relation between the
phase gradient applied to the IRS and the resulting electronic
rotation angle. Section~\ref{sec:optimization} proposes a set of
algorithms for SRE optimization, while Section~\ref{sec:sensitivity}
discusses the sensitivity of such algorithms to network
parameters. Finally, Section~\ref{sec:results} provides a set of
numerical results obtained by applying the proposed algorithms to a
realistic environment. Conclusions are drawn in
Section~\ref{sec:conclusions}.

\subsection{Mathematical notation}
Boldface uppercase and lowercase letters denote matrices and
vectors, respectively. $\Id_k$ is the $k \times k$ identity matrix and
$\onev_k$ is the all-1 column vector of length $k$.  The conjugate
transpose of matrix $\Am$ is denoted by $\Am\Herm$, while
$[\Am]_{i,j}$ is its $(i,j)$-th element.  $\Am^+$ and
$\|\Am\|_{\rm F}$ refer, respectively, to the Moore-Penrose
pseudo-inverse and the Frobenius norm of $\Am$.  The notation
$\Am=\diag(\av)$ specifies that the entries of the vector $\av$ are
the elements of the diagonal matrix $\Am$. Symbols
$\otimes$ and $\EE[\cdot]$ denote the Kronecker product and the average
operator, respectively.  Finally, we 
define the norm-1 length-$M$ column vector $\sv(\Delta, M, \beta)$, whose
$m$-th entry is 
\begin{equation} [\sv(\Delta, M, \beta)]_m =
\frac{1}{\sqrt{M}}\ee^{\jj \pi \Delta(M-1) \sin \beta} \ee^{-\jj 2 \pi
  \Delta(m-1) \sin \beta}
\label{eq:notation}
\end{equation}
for $m=0,\ldots,M-1$. This vector represents the (normalized) spatial signature of a uniform
linear array (ULA) composed of $M$ elements spaced by $\Delta$
wavelengths as observed from the angle $\beta$, measured from the normal to
the ULA.

\section{Communication model\label{sec:model}}
We consider the downlink of a wireless network operating in the
sub-THz band. The network is composed of a BS, which transmits $K$
data streams to $K$ users (UEs). We assume that the UEs
  are not in LoS with the BS. Nevertheless, they can receive the BS
  signal through a NLoS reflection provided by a wall and through a
  set of $N$ IRS, as depicted in Fig.~\ref{fig:map}. In order to
simplify the discussion and the mathematical description of the
system, we assume that transceivers, IRSs and reflectors have the same
height above ground (2D model), so that all the relevant angles lie on
the azimuth plane\footnote{The extension to a 3D scenario is
  straightforward and does not add significant insight on the system
  behavior.}.  In the following we provide a detailed description of
the BS, the IRSs and the UEs, as well as of the channel model.

\subsection{Base Station}
The BS is equipped with an ULA composed of $M_1$ isotropic antennas,
spaced by $\Delta_1$ wavelengths. Thus, the transmitted signal,
$\tv$, can be represented by the $M_1\times 1$ vector
\begin{equation}
  \label{eq:t}
  \tv = \Gammam \xv\,,
\end{equation}
where $\xv=[x_1,\ldots, x_K]\Tran$ is the vector of transmitted
symbols for the $K$ users, supposed to have zero mean and covariance
$\EE[\xv\xv\Herm]=\Id_K$, and $\Gammam$ is a precoding matrix. We 
assume that the transmit power cannot
exceed $\Pc_t$, i.e.,
\begin{equation}
  \EE[\|\tv\|_2^2] =\EE_{\xv}\left[\|\Gammam \xv\|_2^2 \right] =\| \Gammam \|^2_{\rm F} \le \Pc_t\,.
  \label{eq:transmit_power}
\end{equation}

\subsection{User Equipments}
At the receiver side, we assume that the UEs are equipped with ULAs as
well. Specifically, each UE ULA is composed of $M_2$ isotropic
antennas spaced by $\Delta_2$ wavelengths.  Hence, the signal received
by the $k$-th UE can be described by the $M_2\times 1$ vector
\begin{equation}
  \rv_k = \widetilde{\Hm}_k\tv + \zv_k\,,
  \label{eq:rv}
\end{equation}
where $\widetilde{\Hm}_k$ is the $M_2\times M_1$ channel matrix
connecting the BS to the $k$-th UE, $\tv$ is given by~\eqref{eq:t},
and $\zv_k$ is a vector of i.i.d.  complex, circularly symmetric
Gaussian noise random variables with zero mean and variance
$\sigma^2$. We assume that the UE ULAs can only perform analog
beamforming, since they are supposed to have limited hardware
complexity. Thus, the $k$-th UE applies to $\rv_k$
the beamforming vector $\fv_k=\sv(\Delta_2, M_2, \alpha_k)$
and forms the output
\begin{equation}
  \label{eq:yk}
  y_k = \fv_k\Herm \rv_k = \fv_k\Herm\widetilde{\Hm}_k\tv + \eta_k\,, 
\end{equation}
where $\eta_k=\fv_k\Herm \zv_k$ is a complex Gaussian random variable
with zero mean and variance $\sigma^2$, and $\alpha_k$ is the
direction of the $k$-th UE beam, measured from the direction normal
to the UE ULA. In conclusion, the signal received by the $K$ UEs can
be described by the vector
\begin{equation}
  \label{eq:y}
  \yv = \left[\begin{array}{c} y_1\\ \vdots \\y_K\end{array} \right] = \underbrace{\left[\begin{array}{c}\fv_1\Herm\widetilde{\Hm}_1 \\ \vdots \\ \fv_K\Herm\widetilde{\Hm}_K\end{array} \right]}_{\widetilde{\Hm}}\tv +\etav\,,
\end{equation}
where $\widetilde{\Hm}$ is the overall channel matrix and $\etav=[\eta_1,\ldots,\eta_K]\Tran$. 

\subsection{Sub-THz communication channel}

\subsubsection{Single-hop channel model}

At sub-THz/THz frequencies, the wireless channel has not
  been completely characterized yet. We know that the main components are the LoS and NLoS reflected rays while scattering and
  diffraction provide marginal contribution~\cite{Akyildiz2018,tarboush2021teramimo}.  Moreover,
  the number of multipath (reflected) components is typically very
  small and even reduces to one when large, high-gain antenna arrays
  are employed~\cite{Akyildiz2018}. Furthermore, NLoS paths are
  subject to severe reflection losses and other important aspects such
  as molecular absorption, blockage, and large-scale fading effects
  need to be taken into account.

  Then, a single hop link between any two devices equipped with
  antenna arrays composed of, respectively, $p$ and $q$ elements, 
can be described by the $p\times q$ matrix~\cite{3gppchanmodel,Akyildiz2018}
    \begin{equation}\label{eq:channel_model}
      {\boldsymbol \Hc} = \underbrace{a_0 c_0 \pv_0\qv_0}_{\rm LoS} + \underbrace{\sum_{p=1}^Pa_pc_p \pv_p\qv_p\Herm}_{\rm NLoS}\,, 
    \end{equation}
    where the vectors $\pv_0=\pv(\varphi_{{\rm r},0})$ and
    $\qv_0 = \qv(\varphi_{{\rm t},0})$ are, respectively, the spatial
    signatures of the receive and transmit array in the direction of
    the LoS path, which is observed at the angles
    $\varphi_{{\rm r},0}$ and $\varphi_{{\rm t},0}$ by the transmitter
    and by the receiver, respectively.  Moreover, $P$ is the number of
    NLoS paths, and the vectors $\pv_p = \pv(\varphi_{{\rm r},p})$ and
    $\qv_p = \qv(\varphi_{{\rm t},p})$, $p=1,\ldots,P$ are the receive
    and transmit array signatures for the $p$-th reflector, which is
    observed at the angles $\varphi_{{\rm t},p}$ and
    $\varphi_{{\rm r},p}$.  The random variables $a_p$ model
    large-scale fading whereas $c_p$ account for the attenuation and
    phase rotation due to propagation. Specifically, we have
    \begin{equation}\label{eq:c_LOS}
      c_p = \rho_p\sqrt{\frac{G_1 A}{4\pi d_p^2}}\ee^{-\jj \frac{2 \pi}{\lambda}d_p }\ee^{-\kappa d_p} = \rho_p\frac{\sqrt{G_1 G}}{4\pi d_p/\lambda}\ee^{-\jj \frac{2 \pi}{\lambda} d_p}\ee^{-\kappa d_p}\,,
    \end{equation}  
    where $\rho_p$ is the complex reflection coefficient ($\rho_0=1$
    for the LoS path), $d_p$ is $p$-th path length and $G_1$ is the
    array gain on one side of the link. The first expressions in~\eqref{eq:c_LOS}
    is employed if the array on the other side is considered in terms
    of its effective area $A$, the second expression is preferred if
    the array is characterized by its gain, $G$. Finally, the
    coefficient $\kappa$ represents the (frequency dependent)
    molecular absorption coefficient \cite{3gppchanmodel},
    \cite{Molecular}.

  \begin{figure}[t]
\centerline{\includegraphics[width=0.4\columnwidth]{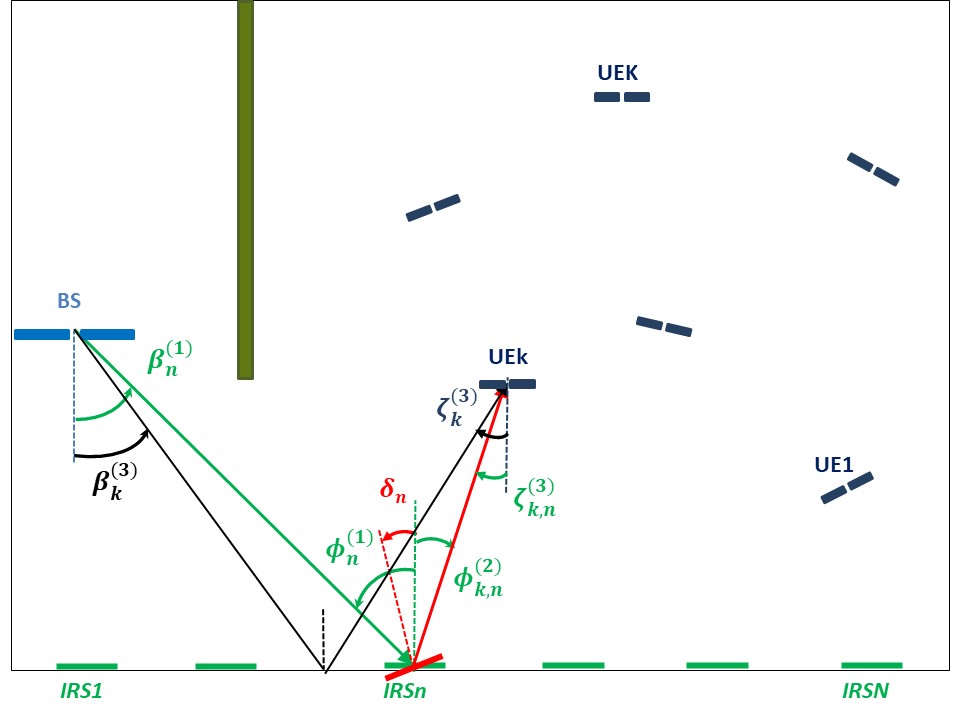}}
\vspace{-2ex}
\caption{An example of SRE where a BS communicates with $K$ UEs by
  exploiting a set of $N$ IRSs deployed along a wall. The direct LoS
  path between BS and UEs is blocked by an obstacle (dark green
  block). Green and red solid lines refer to the LoS paths connecting
  the BS to the $n$-th IRSs and the $n$-th IRS to the $k$-th UE,
  respectively. The black line denotes the NLoS link connecting the BS
  with the $k$-th UE through reflection on a wall (dark thick line at the bottom of the figure).}
\label{fig:map}
\end{figure}

\subsubsection{End-to-end channel model}
In our model the LoS link between the BS and the UEs is blocked by the
presence of an obstacle, represented in Fig.~\ref{fig:map} by the dark green rectangle. However UEs can
receive copies of the BS signal reflected by all IRSs as well as the
one reflected by a wall, represented by the dark thick line at the bottom of Fig.~\ref{fig:map}.
 
Then, the channel matrix $\widetilde{\Hm}_k$ in~\eqref{eq:rv} can be written as
\begin{equation}
  \widetilde{\Hm}_k = \sum_{n=1}^N \Hm^{(2)}_{k,n}\bar{\Thetam}_n\Hm^{(1)}_{n} + \Hm^{(3)}_k\,,\qquad \mbox{where}
\label{eq:Hk}
\end{equation} 
\begin{itemize}
\item $\Hm^{(1)}_{n}$ is the $L_n^2\times M_1$ channel matrix connecting the BS to the $n$-th IRS;
\item $\bar{\Thetam}_n=\Id_{L_n}\otimes \Thetam_n$ is the diagonal
  matrix of the phase shifts introduced by the meta-atoms of
  the $n$-th IRS, where $\Thetam_n=\diag(\ee^{\jj \theta_{n,1,1}}, \ldots,
  \ee^{\jj\theta_{n,L_n,1}})$, $\theta_{n,\ell,\ell'}$ being defined in~\eqref{eq:phase_shift};
\item $\Hm_{k,n}^{(2)}$ is the $M_2\times L_n^2$ channel matrix
  connecting the $n$-th IRS to the $k$-th UE;
\item $\Hm^{(3)}_k$ is the channel matrix connecting the BS to the $k$-th UE, through wall reflection.
\end{itemize}
In our notation, the superscripts $^{(1)}$, $^{(2)}$ and $^{(3)}$
refer to the link connecting the BS to the IRSs, the
link connecting the IRSs to the UEs, and the path connecting BS and UE
through wall reflection, respectively.

Let us first consider the link connecting the BS and the
  $n$-th IRS. Since their position is fixed, we assume that
  they have been conveniently deployed so that they are
  connected by a dominant unfaded LoS link. Therefore, the channel
matrix $\Hm_n^{(1)}$ in~\eqref{eq:Hk} contains only the LoS component and,
recalling~\eqref{eq:channel_model} it can be written as
\begin{equation}
  \label{eq:chan_1}
  \Hm_{n}^{(1)} = c_n^{(1)} \bar{\uv}_n^{(1)}{\vv_n^{(1)}}\Herm\,, \qquad \mbox{where}
\end{equation}
\begin{itemize}
\item
  $\bar{\uv}_n^{(1)} = \frac{1}{\sqrt{L_n}}\onev_{L_n}\otimes \sv(\Delta, L_n, \phi_n^{(1)})$
  is the spatial signature of the $n$-th IRS and
  $\phi_n^{(1)}$ is the LoS angle of arrival (AoA) of the BS signal at the
  $n$-th IRS, measured with respect to a direction orthogonal to the
  surface (see Fig.~\ref{fig:map});
\item $\vv_n^{(1)}\mathord{=}\sv(\Delta_1, M_1, \beta_n^{(1)})$, where $\beta_n^{(1)}$
  is the LoS angle of departure (AoD) of the signal from the BS towards
  the $n$-th IRS, measured with respect to the direction orthogonal to
  the BS ULA;
\item $c_n^{(1)}\mathord{=}a_n^{(1)}\frac{\sqrt{M_1 A_n\cos
        \phi_n^{(1)}}}{\sqrt{4\pi} d_n^{(1)}} \ee^{-\jj\frac{2\pi}{\lambda} d_n^{(1)}} \ee^{-\kappa d_n^{(1)}}$,
    is the LoS channel gain~\eqref{eq:c_LOS}:
    where
    $A_n$ is given by~\eqref{eq:IRS_area}, and $d_n^{(1)}$ is the
    distance between the center of the BS ULA and the center of the
    $n$-th IRS. We recall that $A_n\cos \phi_n^{(1)}$ is the effective
    area of the IRS as observed from the AoA $\phi_n^{(1)}$. Finally,
    since we consider an unfaded LoS link, we set $a_n^{(1)}=1$.
\end{itemize}

The IRSs--UEs links are, instead, affected by the random
  position of the UEs in an environment prone to shadowing effects and potentially containing objects
  acting as reflectors. Thus, for the link connecting the $k$-th UE with the $n$-th IRS
  we adopt the model in~\eqref{eq:channel_model} accounting for $P\ge 1$ NLoS paths. The corresponding channel matrix, $\Hm_{k,n}^{(2)}$, is given by
  \begin{equation}
  \label{eq:chan_2}
\Hm_{k,n}^{(2)} = \sum_{p=0}^{P}a^{(2)}_{k,n,p}c_{k,n,p}^{(2)}\wv^{(2)}_{k,n,p}  \bar{\uv}^{(2)\, {\mathsf H}}_{k,n,p}\,\quad \mbox{where}
\end{equation}
\begin{itemize}
\item $a_{k,n,p}^{(2)}$ is a random variable with log-normal
  distribution, describing large-scale fading effects on the $p$-th path
  between the $n$-th IRS and the $k$-th UE;
\item
  $\bar{\uv}_{k,n,p}^{(2)} = \frac{1}{\sqrt{L_n}}\onev_{L_n}\otimes
  \sv(\Delta, L_n, \phi_{k,n,p}^{(2)})$, where $\phi_{k,n,p}^{(2)}$ is the
  AoD towards the $p$-th path;
\item $\wv_{k,n,p}^{(2)}= \sv(\Delta_2, M_2, \zeta_{k,n,p}^{(2)})$ is
  the spatial signature of the $k$-th UE ULA, as observed from the direction of the $p$-th path, $\zeta_{k,n,p}^{(2)}$;
\item $c_{k,n,p}^{(2)}= \rho_{\rm IRS} \rho_{k,n,p}\frac{\sqrt{M_2A_n\cos \phi^{(2)}_{k,n,p}}}{\sqrt{4\pi}d_{k,n,p}^{(2)}}
  \ee^{-\jj \frac{2\pi}{\lambda} d_{k,n,p}^{(2)}}\ee^{-\kappa d_{k,n,p}^{(2)}}$
  is the channel gain of the $p$-th path, $\rho_{\rm IRS}$ is the IRS reflection coefficient, $\rho_{k,n,p}$ is the reflection coefficient of the $p$-th reflector, and
  $d_{k,n,p}^{(2)}$ is the distance between the $n$-th
  IRS and the $k$-th UE, through the $p$-th path;
\item for $p=0$, the angles $\phi_{k,n,0}^{(2)}$ and $\zeta_{k,n,0}^{(2)}$ refer to the LoS path.
\end{itemize}
\begin{remark}
  For simplicity we assume the IRS reflection coefficient, $\rho_{\rm
    IRS}$, to be a constant independent on the other system
  parameters. We point out, however, that in some practical settings
  the amplitude response of a meta-atom depends on the applied phase
  shift in~\eqref{eq:phase_shift}, as observed
  in~\cite{Amplitude_phaseshift}.
\end{remark}

Finally, according to the image theorem, the NLoS link connecting the
BS to the $k$-UEs through reflection on the wall (see
Fig.~\ref{fig:map}) can be described by the matrix
\begin{equation}
  \label{eq:Hwall}
\Hm_{k}^{(3)} = a_k^{(3)}c_k^{(3)} \wv^{(3)}_k{\vv_k^{(3)}}\Herm\,,\quad \mbox{where}
\end{equation}
\begin{itemize}
\item $a_k^{(3)}$ is a random variable modeling large scale fading effects;
\item
  $c_{k}^{(3)}= \rho_{\rm wall} \frac{\sqrt{M_2M_1\lambda^2}}{4 \pi
    d_k^{(3)}}\ee^{-\jj \frac{2 \pi}{\lambda} d_k^{(3)}}\ee^{-\kappa
    d_k^{(3)}}$, where $\rho_{\rm wall}$ is the reflection coefficient
  of the wall and $d_k^{(3)}$ is the path length;
\item $\wv_k^{(3)}= \sv(\Delta_2, M_2, \zeta^{(3)}_k)$ and
  $\zeta^{(3)}_k$ is the AoA of the signal reflected by the wall, as
  observed from the $k$-th UE, from the direction orthogonal to the UE
  ULA;
\item $\vv_k^{(3)} = \sv(\Delta_1, M_1, \beta^{(3)}_k)$, and
  $\beta^{(3)}_k$ is the AoD of the signal that is reflected by the
  wall towards the $k$-th UE, as observed by the BS, measured from the
  direction orthogonal to the BS ULA.
\end{itemize}

\subsection{IRS characterization}

{\bf Radiation pattern.} 
  Several power radiation patterns for the IRS elements have been discussed and analyzed in the literature~\cite{W-Tang};
  we here assume that the power of the radiation collected by an IRS
  of area $A$ is proportional to $A\cos \phi^{(1)}$ for $\phi^{(1)}\in [-\pi/2,\pi/2]$, and zero otherwise,
  where $\phi^{(1)}$ is the AoA of the radiation.
  Similarly, we assume that the power radiated by the $n$-th IRS is proportional to $|\rho_{\rm IRS}|^2A\cos \phi^{(2)}$, for  
  $\phi^{(2)}\in [-\pi/2,\pi/2]$ and zero otherwise, where $\phi^{(2)}$ is the AoD of the scattered field and
  $\rho_{\rm IRS}$ is the reflection efficiency of the IRS.
  In practice the terms $A\cos \phi^{(1)}$ and $A\cos \phi^{(2)}$ represent the effective area of the IRS when observed from
  the AoA and AoD, respectively.
  The above expressions take into account that IRSs typically receive power and radiate it on one side only.
  Note that in general the IRS's received and radiated powers also depend on
  the AoA and AoD of the electromagnetic field measured in the elevation plane.
  However, since we employ a 2D description of the system geometry
  (i.e. we work in the azimuth plane) such dependencies can be
  neglected.

  We also assume that the intensity of the scattered electromagnetic
  field decays with the inverse of the distance, and that the IRS are
  uniformly illuminated by the BS; our model is intended to hold in
  the far-field regime~\cite{DiRenzo2, Larsson_2020} and when the {\em
    angular aperture} of the IRS, as observed from the BS, is small
  when compared to the beamwidth of the BS signal.

  {\bf IRS size and phase-shift properties.}
   In our model the $n$-th IRS, $n=1,\ldots,N$, has square
  shape and is composed of $L_n^2$ meta-atoms~\cite{Liaskos}, arranged in a $L_n\times L_n$
square grid, of area
\begin{equation}\label{eq:IRS_area}
  A_n=L_n^2\Delta^2\lambda^2\,,
\end{equation} 
where $\lambda$ is the signal wavelength and $\Delta$
is the meta-atom side length, normalized to $\lambda$.

The meta-atom at position $(\ell,\ell')$ in the $n$-th surface,
$\ell,\ell'=1,\ldots,L_n$, applies a phase shift
$\theta_{n,\ell,\ell'}$ to the signal impinging on it.  We
  here assume that such phase shifts can take any value in $[0,2\pi)$,
  i.e. the IRS elements behave as continuous phase
  shifters. However, practical implementations restrict the possible
  phase shifts to a discrete set, whose cardinality, $2^b$, depends on
  the number of control bits, $b$, per IRS element. It has been shown
  that phase shifters with at least 3 control bits entail small
  performance degradation with respect to continuous phase
  shifters~\cite{discrete_phase,Discrete_phase_shift} and achieve 
    close-to-optimal performance.

Many works assume a rich scattering channel and
thus optimize the system performance by
jointly searching for the appropriate values of each of the phase shifts
$\theta_{n,\ell,\ell'}$~\cite{Full-duplex,multicell, Wu-Zhang,
  HongbinLi-2,HongbinLi-1}.
However, as discussed in Section~\ref{sec:intro}, since we aim at simplicity and we consider a channel
characterized by extreme sparsity and negligible scattering and diffraction effects, we 
assume that phase shifts of the $n$-th IRS are
related to each other according to the linear
equation~\cite{Wymeersch, noiICC2020,scattermimo}
 \begin{equation}\label{eq:phase_shift} \theta_{n,\ell,\ell'}
  = 2\pi g_n \Delta\left(\ell-1-\frac{L_n-1}{2}\right) + \psi_n\,,
\end{equation}
for $\ell'=1\ldots,L_n$.  By virtue of~\eqref{eq:phase_shift}, the
$n$-th IRS is able to steer the impinging signal and beam it to an
arbitrary direction (depending on the parameter $g_n$, which is
proportional to the phase gradient) as well as to apply an arbitrary
phase shift, $\psi_n$, to the reflected signal. Note
that~\eqref{eq:phase_shift} allows to characterize the IRS by using
only two parameters, i.e., $g_n$ and $\psi_n$, regardless of the
number of meta-atoms, $L_n^2$. We point out that, although such IRS model is known to
be optimal for a single-user system in pure LoS condition, it allows
for a simple, practical and efficient network configuration.
Nevertheless, in the following we will show that under particular
assumptions, it can be optimal also in a specific multi-user environment. The derivation of a
closed form expression of the optimal solution in a general multi-user
scenario remains an open problem.

\section{Asymptotic expression of the channel matrix and signal precoding\label{sec:asymptotic}}
The overall channel matrix $\widetilde{\Hm}$ in~\eqref{eq:yk} can be
written in a more tractable form by letting the number of meta-atoms
contained in each IRS tend to infinity, while keeping constant the
surface area.  This is a reasonable assumption since the number of
meta-atoms in an IRS is usually large and the normalized meta-atom
side length, $\Delta$, is typically very small.

\begin{proposition} \label{prop:1}
  As $L_1,\ldots L_N\to\infty$ while the IRS areas $A_n$ remain
  constant, the matrix $\widetilde{\Hm}$ tends to matrix $\Hm$, given
  by
\begin{equation}
  \Hm = \lim_{L_1,\ldots L_N\to\infty}\widetilde{\Hm}= \Mm \Psim {\Vm^{(1)}}\Herm + \Tm {\Vm^{(3)}}\Herm\,,
  \label{eq:H_mat}
\end{equation}
where $\Psim=\diag\left(\ee^{\jj \psi_1},\ldots,\ee^{\jj
    \psi_N}\right)$, $\Vm^{(1)} = [\vv_1^{(1)}, \ldots,\vv_N^{(1)}]$, $\Vm^{(3)} = [\vv_1^{(3)}, \ldots,\vv_K^{(3)}]$,  $\Mm$ is a
$K\times N$ matrix whose elements are given by

\begin{equation}\label{eq:mkn}
  [\Mm]_{k,n}\mathord{=} c_n^{(1)}a_{k,n,p}^{(2)}c_{k,n,p}^{(2)} \sum_{p=0}^Pb_{k,n,p}  \sinc\left(\frac{\sqrt{A_n}}{\lambda}s_{k,n,p} \right)\,,
  \end{equation}
$\Tm$ is a diagonal matrix whose $k$-th diagonal element is given by
\begin{equation}\label{eq:T}
[\Tm]_{k,k} = b_ka_k^{(3)}c_k^{(3)}\,,
\end{equation}
having defined
\begin{equation} b_{k,n,p} \triangleq
  \frac{\sinc(\Delta_2M_2(\sin\alpha_k-\sin \zeta_{k,n,p}))}{\sinc(\Delta_2(\sin \alpha_k-\sin \zeta_{k,n,p}))}\,, \qquad b_k \triangleq 
 \frac{\sinc(\Delta_2M_2(\sin\alpha_k-\sin \zeta_k))}{\sinc(\Delta_2(\sin \alpha_k-\sin \zeta_k))}\,,\label{eq:bknp}
\end{equation}
and 
\begin{equation}\label{eq:tkn}
      s_{k,n,p}=\sin \phi_n^{(1)}-\sin \phi_{k,n,p}^{(2)}-g_n\,.
    \end{equation}
  \end{proposition}
  
\begin{IEEEproof}
  See Appendix~\ref{app:A}.
\end{IEEEproof}

We make the following remarks about~\eqref{eq:mkn}:
\begin{itemize}
\item the term $s_{k,n,p}$ is related to the
    misalignment of the $k$-th UE w.r.t. the beam generated by the
    $n$-th IRS and received through the $p$-th reflector. We recall
    that $p=0$ corresponds to the LoS component between the $n$-th IRS
    and the $k$-th UE. Also, the direction of maximum radiation (for
    the $p$-th path) corresponds to $s_{k,n,p}=0$;
\item as it can be expected,~\eqref{eq:mkn} is similar to the bistatic
  radar equation when a tilted flat plate is considered as target;
\item each IRS generates a beam whose width is proportional to
  $\lambda/\sqrt{A_n}$.  Therefore, larger surfaces generate narrower
  beams;
\item the channel gain $|[\Mm]_{k,n}|^2$ is proportional to the square
  of the product $c_n^{(1)}c_{k,n,p}^{(2)}$ which, in turn, is
  proportional to the square of the IRS area, $A_n^2$; as observed
  in~\cite{Zhang2019}, such squared gain shows that the IRSs
  achieve at the same time a beamforming gain and an aperture gain, both proportional
  to $A_n$.
\item the term $\sinc\left(\frac{\sqrt{A_n}}{\lambda}s_{k,n,p} \right)$ in~\eqref{eq:mkn},
  is proportional to the radiation pattern of a continuous metal plate with anomalous reflection properties, as 
  observed in~\cite{Larsson_2020}.
\end{itemize}

In the following, we will replace the matrix $\widetilde{\Hm}$ with
its asymptotic expression $\Hm$ given in~\eqref{eq:H_mat}, so that, by recalling~\eqref{eq:t} the
received signal takes the form
\begin{equation}
  \label{eq:r}   \yv=\Hm \Gammam \xv + \etav\,.
\end{equation}
The precoding matrix $\Gammam$ should be designed so as to adapt the
transmitted signal to the propagation environment.  Several choices
are possible: for example it can be designed to maximize the SINR at
the receivers or to null out the interference among UEs, at a price of
a reduction of SINR. In this work we consider zero-forcing (ZF)
precoding, similarly to what done in~\cite{Huang2019}. Specifically,
we will assume in the following that $\min (M_{1},N) \geq K$. Under
this hypothesis, we can choose $\Gammam$ to satisfy
  \begin{equation}
   \Hm\Gammam = a\Qm^{1/2}\,,\label{eq:HGammaP}
  \end{equation}
  where $\Qm$ is a diagonal matrix and $a$ is a coefficient.
  Indeed, by  substituting~\eqref{eq:HGammaP} in~\eqref{eq:r} we observe that the
  effect of the precoder is to diagonalize the end-to-end channel
  matrix and, by consequence, make the UEs' channels orthogonal.
  By solving~\eqref{eq:HGammaP} for $\Gammam$, the precoder can be written as
  \begin{equation}\label{eq:precoder}
    \Gammam = a\Hm^+\Qm^{1/2}\,,
  \end{equation}
  where $\Hm^{+}=\Hm\Herm\left(\Hm\Hm\Herm\right)^{-1}$ is the
pseudo-inverse of
$\Hm$ and $a = \frac{\sqrt{\Pc_t}}{\|  \Hm^{+}\Qm^{1/2} \|_{\rm F}}$  
in order to meet the transmit power constraint in~\eqref{eq:transmit_power}. 
With this precoder choice, the received SNR at the  $k$-th
  UE is given by
  \begin{equation}\label{eq:SNR}
    {\rm SNR}_k = \frac{\Pc_t q_k}{\sigma^2\|  \Hm^{+}\Qm^{1/2} \|^2_{\rm F}}\,,
  \end{equation}
  where $q_k$ is the $k$-th diagonal element of $\Qm$.  The SNR
  in~\eqref{eq:SNR} corresponds to a spectral efficiency per user of
  \begin{equation}
    \label{eq:sp_eff}
    R_k = \log_2 \left(1 + \mathrm{SNR}_k\right)\,,
  \end{equation}
  expressed in bit/s/Hz.  Note that, by varying $q_k$ it is possible
  to provide the UEs with different quality of service, i.e.,
  different values of $R_k$. In the special case $\Qm=\Id$, all users
  achieve the same spectral efficiency.  While the expression for
  $\Gammam$ in~\eqref{eq:precoder} is suboptimal in terms of
  achievable rate, it has the advantage of completely removing
  interference among streams at the UEs and, more importantly, allows
  for a relatively simple optimization of the SNR received by the
  UEs, as shown in Section~\ref{sec:optimization}.

\section{Electronic rotation of the IRSs\label{sec:rotation}}
The macroscopic effect of the phase gradient applied to the IRS
meta-atoms is to \emph{electronically rotate} the IRS with respect to
its physical orientation, according to the generalized Snell's law. By
electronic rotation, the beam generated by the IRS can be steered to
point to an arbitrary direction.
The angle of electronic rotation of the $n$-th IRS, denoted by $\delta_n$ (see Fig.~\ref{fig:map} for details), only depends 
on the gradient of the phase shift in~\eqref{eq:phase_shift}, i.e. on the parameter $g_n$.
In order to map the parameter $g_n$ into the corresponding rotation
angle of the IRS, we make the key observation that the term
$s_{k,n,p}$ in~\eqref{eq:tkn} can be rewritten as 
\begin{equation}\label{eq:tkn2}
   s_{k,n,p} = \sin \left(\phi_n^{(1)}-2\delta_n\right) -\sin\left(\phi_{k,n,p}^{(2)}\right)\,,
\end{equation}
where we recall that the angles $\phi_n^{(1)}$ and
$\phi_{k,n,p}^{(2)}$ represent the AoA of the signal received at the
$n$-th IRS and the AoD from the $n$-th IRS towards the $p$-th path on
the link connecting $k$-UE, respectively,
measured in the azimuth plane and with respect to a direction
orthogonal to the surface.
Note that the angles $\phi_n^{(1)}-\delta_n$ and
$\phi_{k,n,p}^{(2)}+\delta_n$ are the above mentioned AoA and the AoD,
respectively, {\em as seen from the electronically rotated surface}.
The relation between the phase gradient, $g_n$, and the rotation
angle, $\delta_n$, can be immediately derived by
equating~\eqref{eq:tkn} and~\eqref{eq:tkn2}, and is given by $g_n =\sin \left(\phi_n^{(1)}\right) -\sin \left(\phi_n^{(1)}-2\delta_n\right)$.
In the following, we will drop the expression for $s_{k,n,p}$ in~\eqref{eq:tkn} in favor
of~\eqref{eq:tkn2}, since the angle $\delta_n$ has a clearer geometric
interpretation than $g_n$. Thus, if we want to point the beam
generated by the $n$-th surface in the generic direction $\phi^{(2)}$, we must set the
rotation angle $\delta_n$ in~\eqref{eq:tkn2}, so as to have
$\sin \left(\phi_n^{(1)}-2\delta_n\right) -\sin\left(\phi^{(2)}\right)=0$.

\section{Smart Radio Environment Optimization\label{sec:optimization}}
We now aim at maximizing the SNR in~\eqref{eq:SNR},
over the variables $\deltav=[\delta_1,\ldots,\delta_N]\Tran$,
  $\psiv=[\psi_1,\ldots,\psi_N]\Tran$, and $\alphav=[\alpha_1,\ldots,\alpha_K]\Tran$, characterizing the rotations
  and phase shifts of the IRSs, and the direction of the beams generated by the UE
  ULAs, respectively. In practice, for a given matrix $\Qm$, in view of~\eqref{eq:SNR}, we face
  the following optimization problem
  \begin{equation}
    {\rm SNR}_k^{\rm opt} =  \max_{\xiv}{\rm SNR}_k =\frac{\Pc_tq_k}{\sigma^2 \displaystyle\min_{\xiv}\|  \Hm^{+}\Qm^{1/2} \|^2_{\rm F}}\,.\label{eq:optimization_problem}
  \end{equation}
where $\xiv=[\deltav\Tran, \psiv\Tran, \alphav\Tran]\Tran\in [0,2\pi]^{2N+K}$.
As shown in~\eqref{eq:optimization_problem}, maximizing the SNR is equivalent to minimizing the term
  $\| \Hm^{+}\Qm^{1/2} \|_{\rm F}$ which, in general, is not a convex
  function of $\xiv$.  

  To solve this problem, we first propose a semi-analytic approach based on the
  Newton-Raphson method, outlined in Sec.~\ref{sec:NR} and, then, we propose a heuristic
  optimization algorithm, described in Sec.~\ref{sec:heu}.
  
\subsection{Newton-Raphson SRE optimization\label{sec:NR}}
The optimal value for $\xiv$ solving~\eqref{eq:optimization_problem}, in the following
denoted by $\xiv_{\mathrm{opt}}$, is given by
\begin{equation}\label{eq:optimize}
  \xiv_{\mathrm{opt}} = \arg \min_{\xiv \in [0, 2\pi]^{2N+K}} \| \Hm^{+}\Qm^{1/2} \|_{\rm F}^2
  = \arg \min_{\xiv \in [0, 2\pi]^{2N+K}} \trace\left\{(\Hm\Hm\Herm)^{-1}\Qm\right\}\,.
\end{equation}
Note that $\Hm$ is defined by~\eqref{eq:H_mat},~\eqref{eq:mkn},~\eqref{eq:T},
and~\eqref{eq:tkn2}. In particular $\Hm$ depends on $\deltav$ only through matrix $\Mm$, on $ \alphav$ through matrices $\Mm$ and $\Tm$, while it depends on $\psiv$ only through matrix $\Psim$. Finally, matrices $\Vm^{(1)}$ and $\Vm^{(3)}$  are
constant, given the geometry of the system.
Let $f(\xiv) = \trace\left\{(\Hm\Hm\Herm)^{-1}\Qm\right\}$.
We can solve 
numerically \eqref{eq:optimize} by the iterative Newton-Raphson
method: given a starting point $\xiv^{(0)}$, the $h$-th estimate of
$\xiv_{\mathrm{opt}}$, $h = 1,2,\dots$ is given by
\begin{equation} \label{eq:newton}
  \xiv^{(h)} = \xiv^{(h-1)} -
\left[\boldsymbol{\Sc} \left( \xiv^{(h-1)} \right) \right]^{-1}
\nabla f \left( \xiv^{(h-1)} \right)\,,
\end{equation}
where $\boldsymbol{\Sc} = \frac{\partial^2 f}{\partial \xiv
  \partial\xiv\Tran }$ is the Hessian matrix of $f(\xiv)$. The
expressions for $\nabla f$ and $\boldsymbol{\Sc}$ can be obtained in
closed form. A detailed derivation is reported in
Appendix~\ref{app:gradient_and_hessian}.

Iterations of the Newton-Raphson algorithm stop when the magnitude of the
increment from one iteration to the next one falls below a
predetermined threshold. Since $ f(\xiv)$ is not convex, several
different starting points need to be taken, and the final
approximation of $\xiv_{\mathrm{opt}}$ is the (local)
minimum point that yields the smallest value of $f(\xiv)$.

\subsection{Heuristic SRE optimization\label{sec:heu}}

Owing to the complexity of the optimization problem defined in
\eqref{eq:optimization_problem}, we propose a simpler heuristic
approach to environment optimization. Specifically, we make the key
observation that, if the IRS area is large enough, its radiation pattern is characterized by a narrow beam and, thus, will
likely serve a single, properly chosen, UE. By restricting our attention to a
solution where each UE is associated with one IRS and the IRSs--UEs channels are dominated by the
  LoS component, as reasonable in sub-THz/THz propagation, we are able to solve a substantially simpler
problem, at a modest cost in terms of distance from~\eqref{eq:optimization_problem}, as shown in
Section~\ref{sec:results}. Formally, this can be done by defining the
map
\begin{equation} \Mc: \{1,\ldots K \} \to \{1,\ldots,N \}\,, \label{eq:map}
\end{equation}
which associates UE $k$ to IRS $\Mc(k)$.  In practice, this means that
IRS $\Mc(k)$ should be electronically rotated so as to point its beam in the
direction of the UE $k$ and, symmetrically, the UE $k$ steers the beam generated by its ULA so that it
points towards the IRS $\Mc(k)$.  This criterion is particularly
suited when the surfaces are sufficiently large and the UE ULAs have
enough antennas so that the beam generated by IRS $\Mc(k)$ reaches UE
$k$ without interfering with other UEs.

So, under heuristic optimization and given the map $\Mc$, the
electronic rotation angle of the IRS $\Mc(k)$ is set to \beq
\delta_{\Mc(k)} =
\frac{\phi_{\Mc(k)}^{(1)}-\phi_{k,\Mc(k),0}^{(2)}}{2}\,, \eeq which
yields $s_{k,\Mc(k),0} = 0$ in~\eqref{eq:tkn2}, and on the UE side the
beam direction is set to $\alpha_k = \zeta_{k,\Mc(k)}$,
so that $b_{k,\Mc(k),0} = 1$ in~\eqref{eq:bknp}. Regarding phase
shifts, we set the value of $\psi_{\Mc(k)}$ so that the signals reflected
by the IRS and by the wall reach the UE
with the same phase and, thus, generate constructive interference.

It is worth noting that, if $N > K$, there are $N-K$ IRSs which are
not associated to any user. While in a scenario with small surfaces
and single-antenna UEs, the contribution of these IRSs can be
relevant, in the presence of narrow beams generated by the IRSs and high gain UE arrays, their
effect is substantially negligible. If the $n$-th IRS is not associated to any
UE, the heuristic algorithm conventionally set $\delta_n = 0$ and $\psi_n=0$.

Let $\xiv = \xiv(\Mc)$ be the value of $\xiv$ resulting from a given map $\Mc$.
The proposed heuristic
algorithm then consists in finding the optimal map,
$\Mc_\mathrm{opt}$, satisfying
\begin{equation}
\Mc_\mathrm{opt} = \arg \min_{\Mc} \|\Hm_{\mathrm{heu}}^{+}\Qm^{1/2} \|_{\rm F}\,,\label{eq:heuristic}
\end{equation} 
where $\Hm_{\mathrm{heu}}$ is the channel matrix obtained by
setting $\xiv = \xiv (\Mc)$ in the expression for $\Hm$ in~\eqref{eq:H_mat} . Notice that, since there are
$N! / (N-K)!$ possible maps, an exhaustive search of $\Mc_\mathrm{opt}$ is possible only for
small-size scenarios. 

\section{Sensitivity of SRE optimization to system parameters\label{sec:sensitivity}}
In this section, we analyze the sensitivity of SRE optimization, as described in the previous section, to some system parameters. In particular,
we consider the impact on the received SNR in~\eqref{eq:SNR} of the
size of the ULA arrays at the BS and at the UEs, and of the IRS areas. Moreover, in a properly defined limiting regime, we derive a simplified criterion for optimization.

\subsection{Impact of the number of ULA array elements} \label{sec:BS_ant_asymp}
We first notice that the number of 
elements of the BS ULA, $M_1$, appears
in matrix $\Vm^{(1)}$ whose $n$-th column is given by
$\vv_n^{(1)}= \sv(\Delta_1, M_1, \beta_n^{(1)})$.  Such matrix only
depends on the geometry of the system and is not affected by
electronic IRS rotation or phase shifts.

From \eqref{eq:SNR}, the received SNR depends on $\Vm^{(1)}$ through
${\Vm^{(1)}}\Herm \Vm^{(1)}$, whose $(n,n')$ entry is given by \beq
{\vv_n^{(1)}}\Herm \vv^{(1)}_{n'} = \frac1{M_1} \frac{\sin \left( 2
    \pi \Delta_1 M_1 ( \sin \beta_n^{(1)} - \sin
    \beta_{n'}^{(1)})\right)}{\sin \left( 2 \pi \Delta_1 ( \sin
    \beta_n^{(1)} - \sin \beta_{n'}^{(1)})\right)} \label{eq:vn}\,.\eeq

Now, for a given system geometry, the AoDs $\beta_1^{(1)}, \dots, \beta_n^{(1)}$ are
fixed and we suppose that $\beta_n^{(1)} \neq \beta_{n'}^{(1)}$ for $n \neq
n'$. Thus:
\begin{equation} \left| {\vv_n^{(1)}}\Herm \vv^{(1)}_{n'} \right| \leq \frac1{M_1
  \left| \sin \left( 2 \pi \Delta_1 ( \sin \beta_n^{(1)} - \sin
  \beta_{n'}^{(1)})\right) \right| } \stackrel{M_1 \to \infty}{ \longrightarrow} 0 
\label{eq:limit_eq}\end{equation}
for $n \neq n'$\footnote{The larger $M_1$, the narrower the BS
  transmitted beam. Thus, for a too large value of $M_1$, the
  hypothesis that the transmitted beam uniformly illuminates the IRSs
  does not hold. However, as we will see in Section \ref{sec:results},
  for a realistic scenario, this case does not happen.}. This implies
that
$\lim_{M_1 \rightarrow \infty} {\Vm^{(1)}}\Herm \Vm^{(1)} = \Id_N$.
Similarly, the parameter $M_1$ also appears in $\Vm^{(3)}$. Then,
provided that $\beta_k^{(3)} \neq \beta_{k'}^{(3)}$ for $k \neq k'$,
$\lim_{M_1 \rightarrow \infty} {\Vm^{(3)}}\Herm \Vm^{(3)} =
\Id_K$. Finally, if $\beta_n^{(1)} \neq \beta_{k}^{(3)}$ for every
pair of $k$ and $n$,
$\lim_{M_1 \rightarrow \infty} {\Vm^{(1)}}\Herm \Vm^{(3)} =
\mathbf{0}_{N \times K}$.  As a consequence, when IRSs and UEs are
angularly separated with respect to the BS,
  \begin{eqnarray}
    \|\Hm^{+}\Qm^{1/2} \|_{\rm F}^2
    &=& \trace\left\{(\Hm\Hm\Herm)^{-1}\Qm\right\} 
    \stackrel{M_1\to \infty}{\longrightarrow} \trace\left\{(\Mm\Mm\Herm + \Tm\Tm\Herm)^{-1}\Qm\right\}\,.  \label{eq:limitMinf}
  \end{eqnarray}
  Notice that, in this asymptotic regime, the received SNR becomes independent of the phase shifts $\psiv$. This
  happens because, when $M_1$ gets large, the columns of $\Vm^{(1)}$ become
  orthogonal and it is the precoder that allows to properly set the
  phases of the signals impinging to each IRS. Summarizing, we expect
  that \emph{for increasing transmit array size, the impact of phase shifts, $\Psim$, on
the received SNR  decreases until it becomes
    negligible}.

  \subsection{Impact of IRS areas}
  As already observed, from
  the expression of $[\Mm]_{k,n}$ in \eqref{eq:mkn} it can be seen that the width of the
  beam generated by the $n$-th IRS depends on its area, $A_n$. In particular, for
  $\frac{\sqrt{A_n}}{\lambda} \to \infty$:
  \begin{equation} \label{eq:limitAinf} [\Mm]_{k,n} \to c_n^{(1)}a_{k,n,p}^{(2)}c_{k,n,p}^{(2)} \sum_{p=0}^Pb_{k,n,p}  \delta[s_{k,n,p}]\,,
  \end{equation} 
where $\delta[x] =1$ for $x=0$ and $\delta[x] = 0$ otherwise. Thus, it
turns out that the $n$-th IRS contributes to the signal received by
the $k$-th UE only if it points towards one of the paths characterizing the IRS--UE channel. As a consequence, in the presence of a dominant
LoS path we expect that \emph{IRSs with large area
    (compared to $\lambda^2$) should be rotated so as to point
    in the direction of a given UE}.

\subsection{SRE optimization algorithms in the asymptotic regime \label{sec:heuOpt}}
In the doubly asymptotic regime $M_1,A_n \rightarrow \infty$, SRE optimization becomes easier to state
and to solve. Indeed, thanks to \eqref{eq:limitMinf}, the optimization
problem in~\eqref{eq:optimize} reduces to
\begin{equation}
  \label{eq:optimize_simple}
 \widetilde{ \xiv}_{\mathrm{opt}} = \arg
\min_{\widetilde{ \xiv} \in [0, 2\pi]^{N+K}}
\trace\left\{(\Mm\Mm\Herm+\Tm\Tm\Herm)^{-1}\Qm\right\}\,,
\end{equation}
where $\widetilde{ \xiv}=[\deltav, \alphav]\Tran$.  
  Moreover, if we suppose that UEs and multipath reflectors are
  angularly separated, as observed from the IRSs, then, by
\eqref{eq:limitAinf}, we obtain that $ [\Mm]_{k,n} \neq 0$ if and only
if $n = \Mc(k)$. As a consequence, $\Mm\Mm\Herm$ becomes diagonal and
the heuristic problem becomes \beq \label{eq:optimize_simple2}
\Mc_{\rm opt}^{\infty} = \arg\min_{\Mc}\,\, \omega(\Mc)\,, \eeq where
\beq \label{eq:weights} \omega(\Mc) = \sum_{k = 1}^K \left( \left|
     [\Mm]_{k,\Mc(k)}\right|^2 + \left|
     [\Tm]_{k,k}\right|^2\right)^{-1} q_k\,. \eeq The above is an
     assignment problem whose weights are the sums of the squared
     magnitudes of the entries of $\Mm$ and $\Tm$. The solution can be
     found by resorting to, e.g., the Hungarian
     algorithm~\cite{Hungarian_algorithm}.  It is important to
     highlight that the Hungarian algorithm has very low, namely,
     cubic complexity, hence, it can be
     efficiently used for realistically-sized problem instances.

     We point out that, in a particular setup, the
       solution of~\eqref{eq:optimize_simple2}, together with
       \eqref{eq:phase_shift} is indeed the optimal choice for the
       phase shifts $\theta_{n,\ell,\ell'}$.  To introduce the next
       proposition, we define $\mu_{kn}$ as the magnitude of the
       channel reaching user $k$ through IRS $n$.

\begin{proposition} \label{prop:2} Consider the system in
  \eqref{eq:y}, with $\widetilde{\Hm}_k$ given by
  \eqref{eq:Hk}. Suppose $N = K = 2$, $P = 0$ (i.e., no multipath) and
  no wall reflection. Let the BS have full CSI and perform ZF
  precoding. Finally, we concentrate on the asymptotic scenario in
  which $M_1 \rightarrow \infty$ and $L_n \rightarrow \infty$. Under
  these hypotheses, if $\mu_{11}>\mu_{21}$ and $\mu_{22}>\mu_{12}$, or, conversely, $\mu_{11}<\mu_{21}$ and $\mu_{22}<\mu_{12}$, the solution of \eqref{eq:optimize_simple2} (with
  the phase shifts in \eqref{eq:phase_shift}) is optimal in the sense
  that it minimizes $\| \widetilde{ \Hm}^{+} \|^2_{\rm F}$
  (see \eqref{eq:optimization_problem}).
\end{proposition}

\begin{IEEEproof}
See Appendix \ref{app:C}.
\end{IEEEproof}

In the next section, we will show when, in a realistic scenario,
conditions for the asymptotic regime are met. In such conditions, the
heuristic algorithm in its simplified version
\eqref{eq:optimize_simple2}-\eqref{eq:weights} represents a feasible
way of SRE optimization. We observe that the ZF
precoder $\Gammam$ tends to be a simple equalizing beamformer in the
asymptotic regime, since the channels corresponding to the $K$ users
become orthogonal by themselves, and $\Gammam$ tends to split the
power among the different channels in order to meet the relative
quality of service dictated by matrix $\Qm$.

\section{Performance comparison of the proposed optimization
  algorithms\label{sec:results}}
\begin{figure}
  \centerline{\includegraphics[width=0.5\columnwidth]{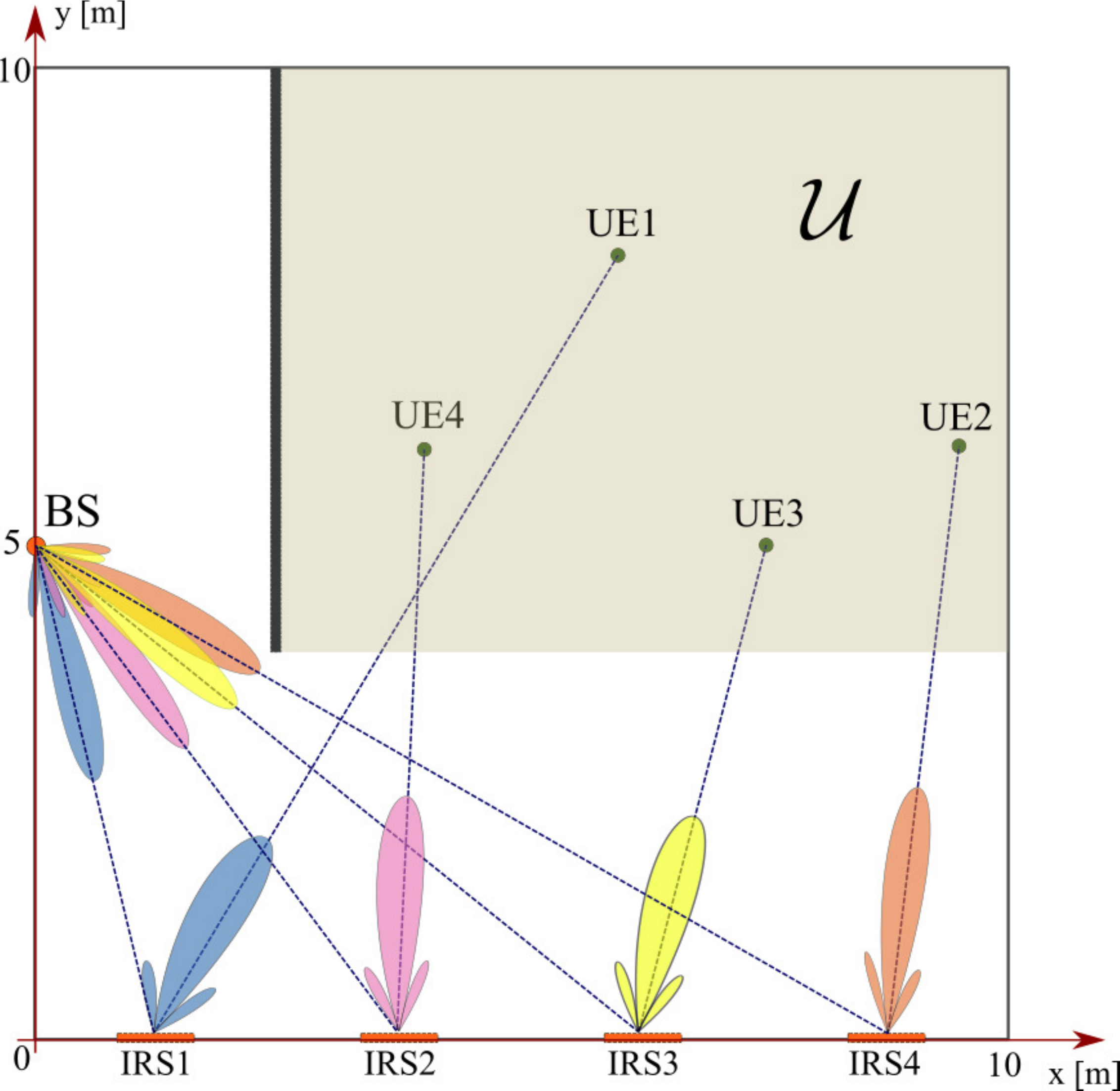}}
  \vspace{-2ex}
  \caption{An example of the system geometry considered in
    Section~\ref{sec:results} for $N=K=4$. The IRSs are equally
    spaced, and the UE are uniformly distributed in the area
    $\Uc$. The plasterboard reflecting wall coincides with the $x$ axis.}\label{fig:figure_1}
\end{figure}

We now assess the performance of the optimization algorithms proposed
in Section~\ref{sec:optimization} and show the influence of the
system parameters on the SNR and on the achievable rate at the UEs. To this purpose we
consider the test scenario in Fig.~\ref{fig:figure_1}, depicting an
area of 100\,m$^2$, whose vertices, are the points $(0,0)$, $(10,0)$,
$(0,10)$ and $(10,10)$ (all coordinates expressed in meters).  The BS
is located at $(0,5)$ and the $N$ IRSs, denoted by the labels IRS1,
$\ldots$, IRS$N$, have area $A_n=A$ and are equally spaced along a
wall coinciding with the $x$ axis.  The positions of the $K$ UEs, denoted by the labels
UE1,$\ldots$,UE$K$, are random variables, uniformly distributed in the rectangle $\Uc$
whose vertices are $(2.5,4)$, $(10,4)$, $(10,10)$ and $(2.5,10)$.

The signal transmitted by the BS has bandwidth $B=100$\,MHz and
carrier frequency $f_0 = 0.1$\,THz, corresponding to the wavelength
$\lambda = 3$\,mm. At such frequency the attenuation due to molecular
absorption is negligible~\cite{3gppchanmodel}; we therefore set
$\kappa=0$ in~\eqref{eq:c_LOS}.  The BS transmit power is $\Pc_t=1$\,W
and we set $\Qm = \Id$, i.e., all users have the same received
SNR. Finally, the separation of the elements of the BS and UE ULAs is
set to $\Delta_1=\Delta_2 = \lambda/2$.

\begin{figure}[t]
 \centerline{\includegraphics[width=0.5\columnwidth]{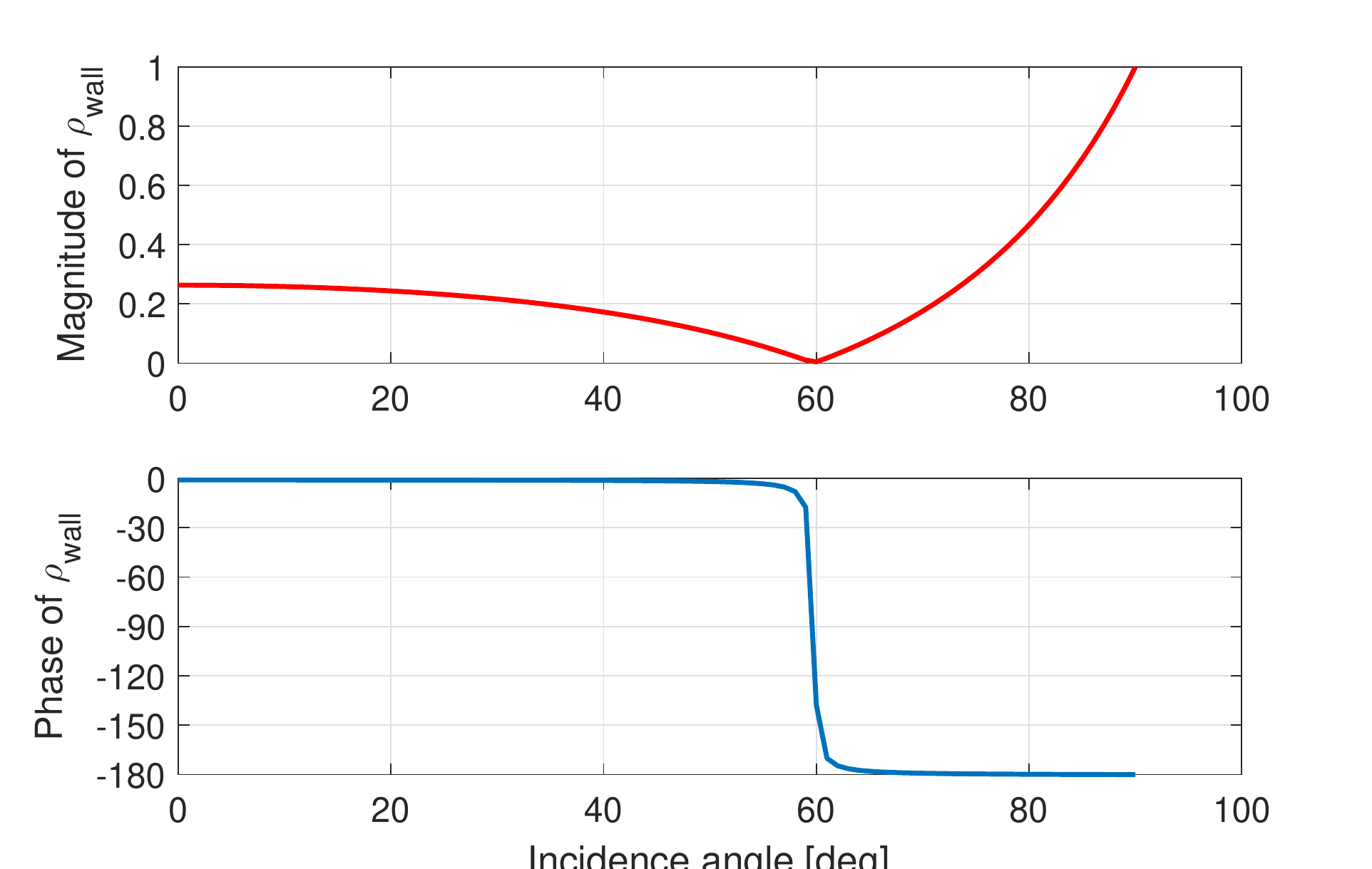}}
 \vspace{-2ex}
\caption{Amplitude (above) and phase (below) of the complex reflection
  coefficient $\rho_{\rm wall}$, for plasterboard panels, plotted
  versus the AoA of the BS signal.  The incidence angle is measured
  from the direction orthogonal to the wall.}\label{fig:wall_reflection}
\end{figure}

The noise power at the receivers is set to $\sigma^2=N_0B$ where
$N_0=-174$~dBm/Hz. We also assume ideal reflection at the IRSs, i.e.,
$\rho_{\rm IRS} = 1$. The wall acting as reflector for the BS signal coincides with the $x$ axis
depicted in Fig.~\ref{fig:figure_1}. We assume it is made of plasterboard whose reflection coefficient,
$\rho_{\rm wall}$, is plotted in Figure~\ref{fig:wall_reflection} versus
the AoA of the BS signal.
The IRS-UE channels are random and affected by multipath and log-normal shadowing. 
Unless otherwise stated, we assume full knowledge of the channel state at the BS.

We evaluate the performance of the following techniques to solve the problem in~\eqref{eq:optimization_problem}:
\begin{itemize}
\item the joint optimization of the IRS electronic rotation angles,
  $\deltav$, of the phase shifts, $\psiv$, and of the UE beam directions, $\alphav$,
  by employing the Newton-Raphson algorithm, in the following referred
  to as ``NRP'';
\item the joint optimization of the IRS electronic rotation
  angles and of the UE beam directions by employing the Newton-Raphson
  algorithm, while setting to zero the IRS phase shifts. This 
  technique, denoted as ``NR'', solves~\eqref{eq:optimization_problem} by imposing
  $\psiv=\zerov$;
\item the heuristic optimization algorithm, ``HOP'' described in
  Section~\ref{sec:heu} and based on the evaluation
  of~\eqref{eq:optimize_simple2}-\eqref{eq:weights}. This algorithm takes
  the phase shifts, $\psiv$, into account for the optimization.
\end{itemize}
Since, in general, the expression of the SNR in~\eqref{eq:SNR} is not
convex in  $\xiv$, for each instance of the system geometry and of the channel,
we perform $U=100$ runs of the ``NR'' and ``NRP'' algorithms, each
characterized by a different, randomly generated, starting point
$\xiv_u^{(0)}$, $u=1,\ldots,U$ and an output ${\rm SNR}_u$.
Then, for each realization of the UE
positions, the SNR provided by the algorithms is given by $\max_u{\rm SNR}_u$. 

The numerical results are organized as follows: in
Section~\ref{sec:CaseStudy}, we show examples of the radiation
patterns emitted by the IRSs and by the BS ULA, while in
Section~\ref{sec:Performance} we compare the performance of the above
optimization techniques in terms of the achieved SNR. Finally, in
Section~\ref{sec:Impact}, we evaluate the impact of the system
parameters on the network throughput.

\subsection{Radiation patterns\label{sec:CaseStudy}} 
We first describe the system behavior in a simple case where we
neglect (i) shadowing effects, (ii) the reflection due to the plasterboard
wall, and (iii) the existence of NLoS paths in the IRS-UE links.  We
also consider a BS ULA with $M_1=32$ elements, $N=4$ IRSs of area
$A=100$\,cm$^2$ and $K=4$ UEs equipped with a single isotropic antenna
($M_2=1$), whose positions are shown in Fig~\ref{fig:figure_1}.
The ``HOP'' algorithm applied to this scenario selects
the IRS-UE assignment depicted in Fig.~\ref{fig:figure_1} by solid
lines. Specifically, the IRSs 1,2,3, and 4 are electronically rotated
so as to point their beams, respectively, towards UEs 1,4,3, and 2.
Referring to~\eqref{eq:map}, this assignment corresponds to the map $\Mc(1)=1$,
$\Mc(2)=4$, $\Mc(3)=3$, and $\Mc(4)=2$.

\begin{figure}
\includegraphics[width=0.5\columnwidth]{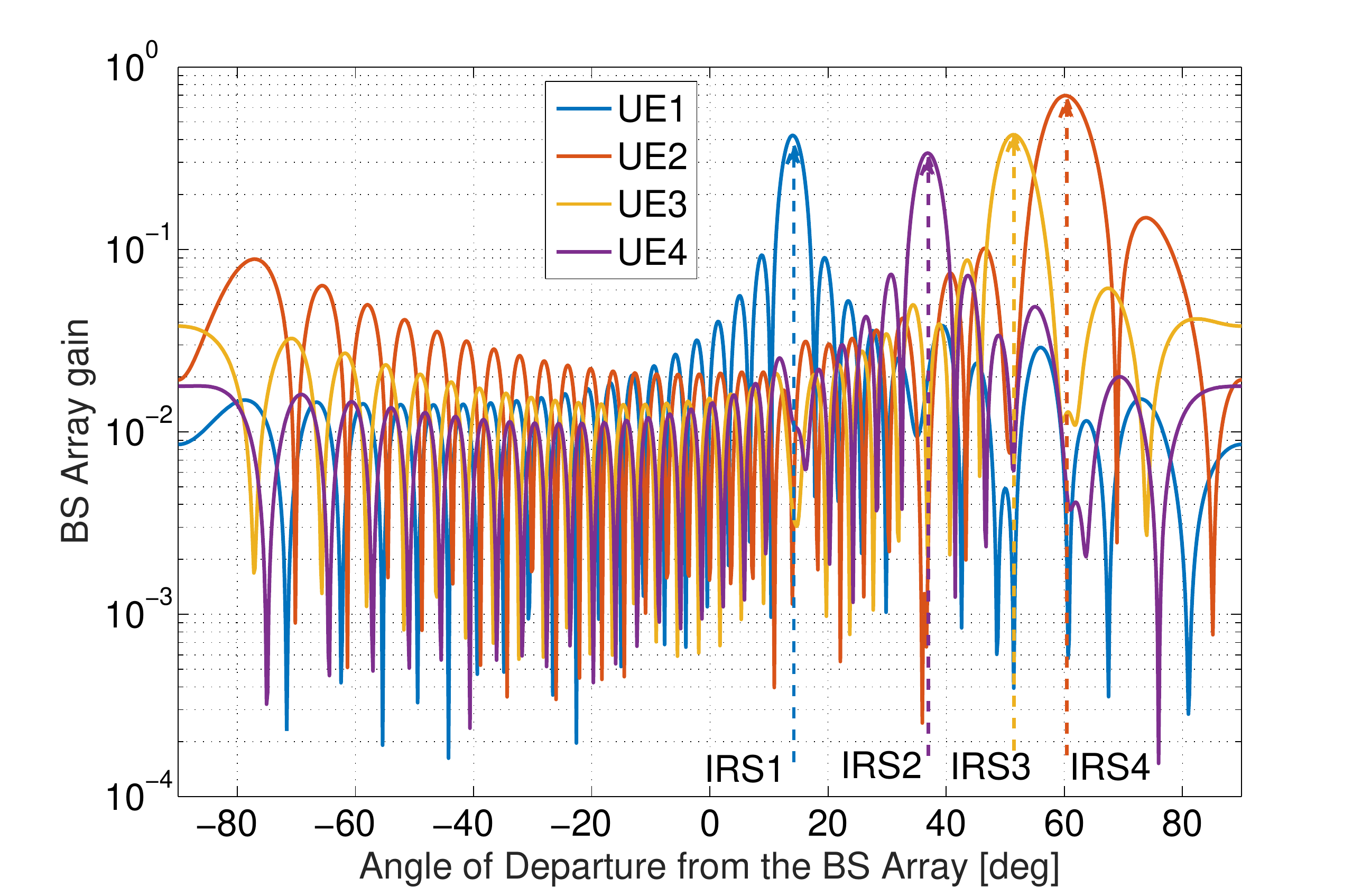}
\includegraphics[width=0.5\columnwidth]{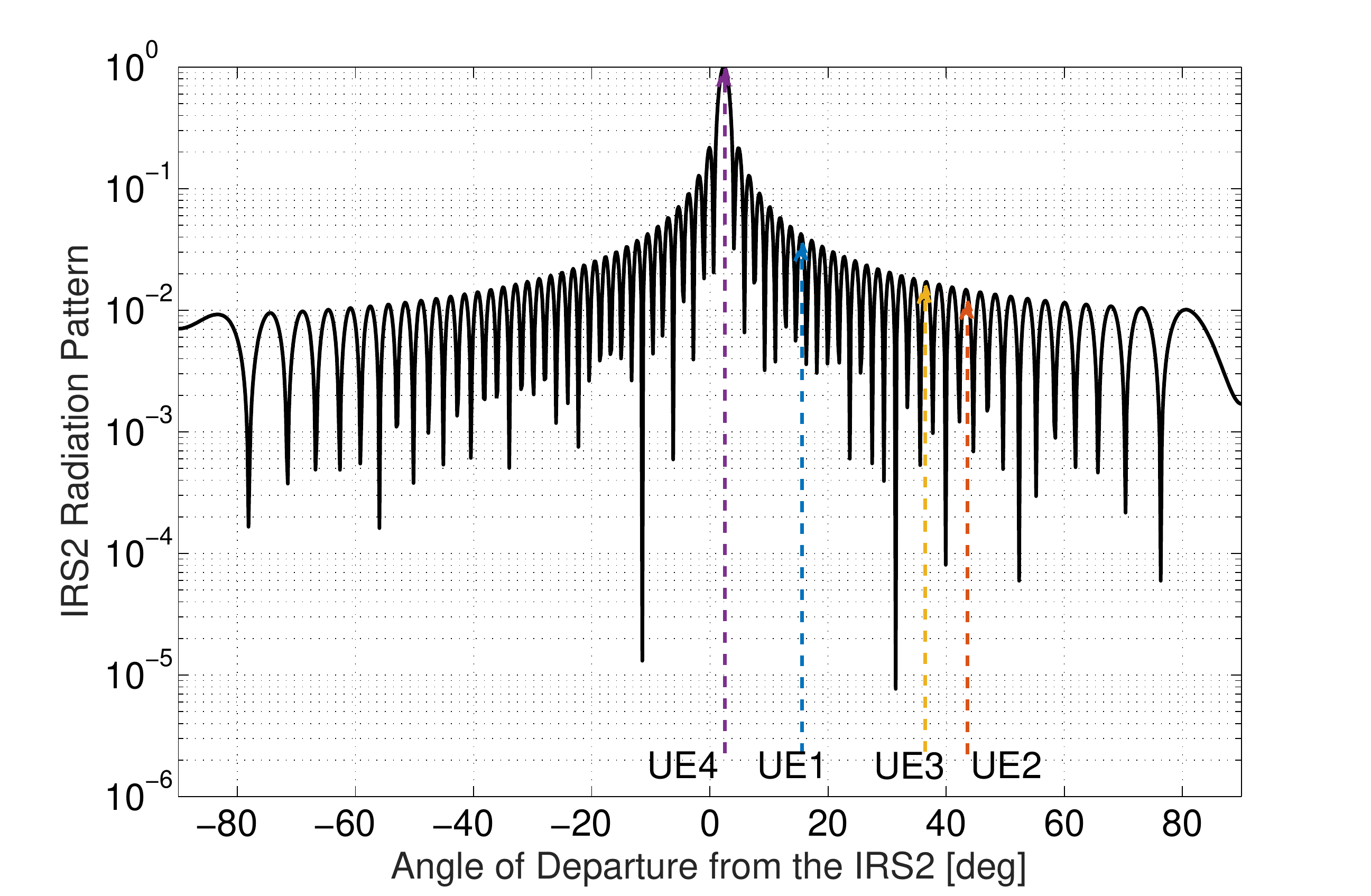}
\vspace{-5ex}
  \caption{(Left) BS array gains $|G_k|^2$, for each of the $K=4$ data
    streams in the scenario depicted in Fig.~\ref{fig:figure_1}.  The
    dashed arrows indicate the directions of the IRSs as observed from
    the BS. (Right) Radiation pattern of IRS2, whose size is $A$ = $100$\,cm$^2$, for
    the IRS-UE assignment as shown in Fig.~\ref{fig:figure_1}. The
    dashed lines indicate the directions of the users as observed by
    the IRS.}\label{fig:Radiation_patterns}
\end{figure}

The BS, thanks to the precoder $\Gammam$, generates $K$ beams, one for each UE.
The radiation pattern of the $k$-th beam as a function of the AoD from the BS ULA, denoted by $\beta$, is given by 
$G_k = \sv(\Delta_1, M_1, \beta)^H \gammav_k$, $k=1,\ldots,K$, where 
$\gammav_k$ is the $k$-th columns of the precoder $\Gammam$. 
To get insight on how the signal energy is distributed among the IRSs, in
Fig.~\ref{fig:Radiation_patterns}(left) we show the array gains
$|G_k|^2$, $k=1,\ldots,4$ plotted versus $\beta$.
As can be observed, the radiation pattern for the data stream intended
for UE1 (blue line) clearly shows a main lobe in the direction of the
IRS1 since such surface steers the signal towards UE1.  Similarly, the
radiation patterns for UE2, UE3 and UE4 show peak values in the
direction of their associated IRS4, IRS3 and IRS2,
respectively. However, note that a fraction of the signal energy
intended for UE1 is also sent to IRSs other than IRS1, due to side
lobes of the radiation pattern.  Similarly, side lobes in the IRS
radiation pattern may generate interference at the UEs as shown by
Fig.~\ref{fig:Radiation_patterns}(right), which reports the radiation
pattern of IRS2. In the figure, the dashed lines indicate the
directions of the users as observed by IRS2. As can be seen, although
the main lobe is directed towards UE4, a side lobe points towards UE1,
albeit with a 15-dB lower gain. However, the resulting interference at
the UEs is canceled out by a proper setting of the IRSs phase
shifts. Thus, the overall user channels are orthogonal, as granted by
the ZF filter. Fig.~\ref{fig:Radiation_patterns}(right) also shows
that the IRS behaves as an ``anomalous'' reflector. Indeed, the AoA of
the BS signal at IRS2 is $36.8^\circ$ while the reflected beam has an AoD of
$2.4^\circ$. As mentioned in Section~\ref{sec:asymptotic}, the area of
an IRS affects the beamwidth of its radiation pattern. For an IRS area
of $100$\, cm$^2$, the first-null beamwidth is about $2^\circ$.  We
point out that the radiation pattern in
Fig.~\ref{fig:Radiation_patterns}(right) has been obtained by assuming
an ideal IRS reflection coefficient, $\rho_{\rm IRS}=1$.  However, if
$\rho_{\rm IRS}$ is phase-shift dependent~\cite{Amplitude_phaseshift}
the radiation pattern might differ from that depicted in the figure;
in particular it could show a lower gain of the main lobe and higher
side lobes.  Furthermore, if the system allows to control the
amplitude response of each meta-atom, the radiation pattern can be
designed e.g. to minimize the side lobes power at a price of a slight
increase of the main lobe beamwidth.

\subsection{Effect of the plasterboard wall\label{sec:Performance}}
We now investigate the performance of the ``NR'' ``NRP'' and ``HOP'' optimization algorithms
and the impact of the system parameters on the received SNR.

\begin{figure}[t]
 \includegraphics[width=0.5\columnwidth]{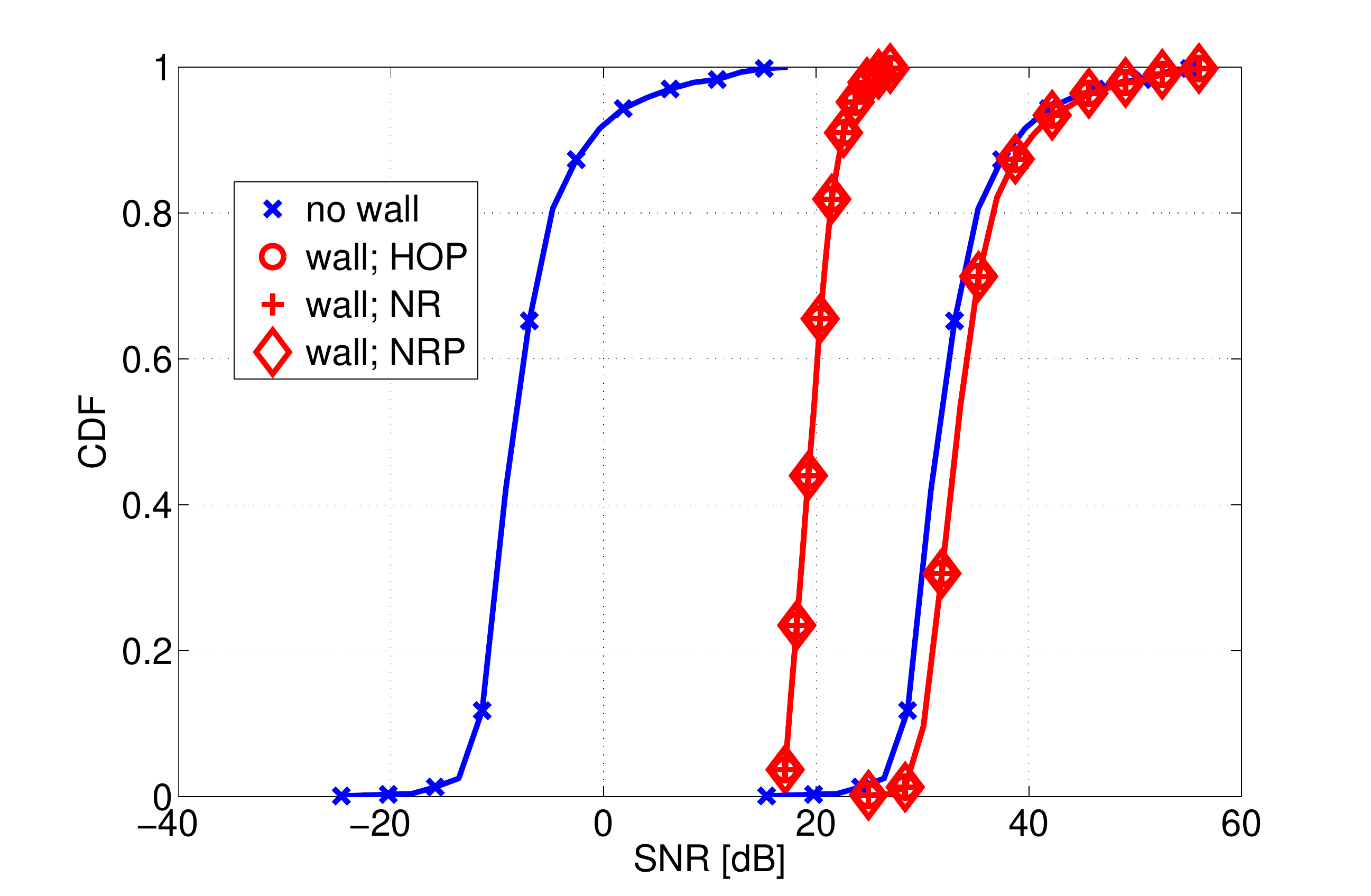}
 \includegraphics[width=0.5\columnwidth]{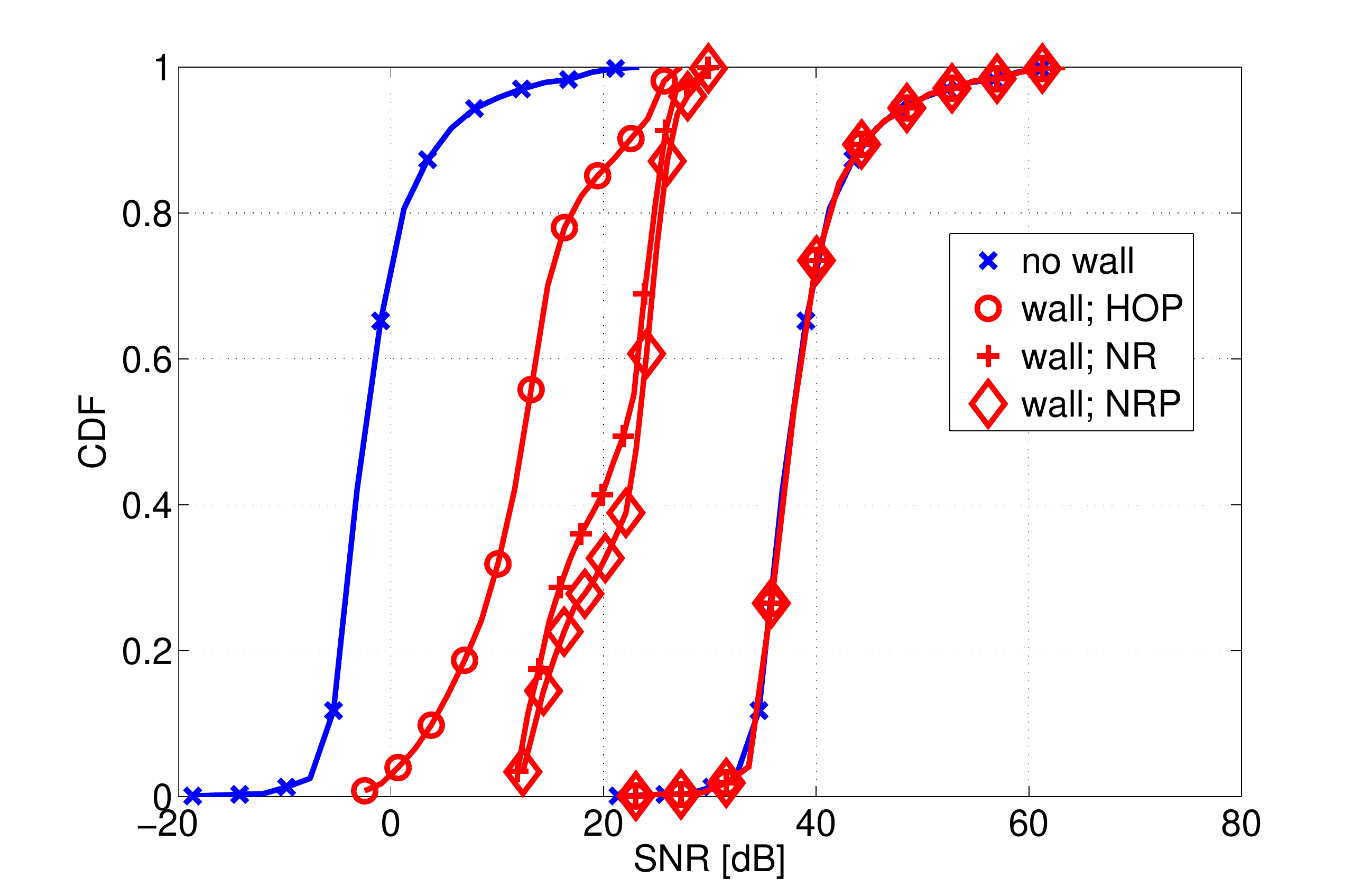}
 \vspace{-5ex}
 \caption{Cdf of the SNR for $N=K=1$, $M_1=4$, $M_2=1$ (left) or
   $M_2=4$ (right) and IRS area $A=[1,100]$\,cm$^2$, in the presence
   or absence of the reflection due to the plasterboard wall. The
   channels are considered singlepath and not affected by shadowing.}
\label{fig:wall_on_off}
\end{figure}

First of all, we measure the effect of the signal reflected by the
plasterboard wall. In Fig.~\ref{fig:wall_on_off}(left) we therefore
consider a single-user scenario ($K=1$), a single IRS ($N=1$), no
multipath on the IRS-UE link ($P=0$), no shadowing, and $M_1=4$ and
$M_2=1$ antenna elements at the BS and UE, respectively.  The
optimization here consists only in the proper choice of the electronic
rotation $\delta_1$ of the IRS1 and of the phase shift $\psi_1$, since
for $M_2=1$ the UE antenna is isotropic and there is no beam direction
to be optimized. Clearly, the choice of $\delta_1$ is trivial, since
the optimum is achieved when the beam generated by the IRS points
towards the UE; in such a case ``NRP'', and ``HOP'' are
expected to provide the same performance.  Then, given the UE position, we choose $\delta_1$ so
as to null $s_{1,1,0}$ in~\eqref{eq:tkn2}.  Moreover, in presence of
the wall reflection, the IRS phase shift $\psi_1$ should be set so as
to ensure constructive interference at the UE.  The cumulative density
function (cdf) of the SNR at the UE, obtained by generating 1000
realizations of the above described scenario, is reported in
Fig.~\ref{fig:wall_on_off}(left) for IRSs of area $A=1$\,cm$^2$ and
$A=100$\,cm$^2$.
For $A=1$\,cm$^2$ the SNR is dominated by the
contribution of the signal reflected by the wall, which yields a gain
of about 25\,dB with respect to the scenario without wall reflection.
Instead, for $A=100$\,cm$^2$, the beneficial contribution of the wall
is limited to about 1.5\,dB. This means that IRS areas should be
accurately designed depending on the number and quality of natural
reflectors in the environment.  IRS with small area provide little
contribution to the received power while larger IRS allow to neglect the
contribution due to natural reflectors.
We also note that, in the absence of wall reflection, by increasing the
IRS area from 1\,cm$^2$ to 100\,cm$^2$ we obtain 40\,dB improvement in
the SNR. This is expected since, from~\eqref{eq:mkn} and as observed
in~\cite{Zhang2019}, the SNR depends on $A^2$.

In Fig.~\ref{fig:wall_on_off}(right), we consider the same setting as in Fig.~\ref{fig:wall_on_off}(left) but 
$M_2=4$ elements. In this case, the received SNR also depends on the
direction, $\alpha_1$, of the beam generated by the UE ULA and,
therefore, the SNR optimization in~\eqref{eq:optimization_problem} is
not as trivial as before, and the ``NR'', ``NRP'', and ``HOP'' algorithms
provide different performance. Specifically, while all algorithms agree that
the IRS should point its beam towards the UE, they return divergent
choices for the angle $\alpha_1$. In particular, 
\begin{itemize}
  \item the ``NRP'' algorithm rotates the UE beam so as to maximize the
    received energy at the UE. The optimal direction is, in general,
    in between the directions of the beams reflected by the IRS and by
    the wall. Also, ``NRP'' adjusts the IRS phase shift, so as to create
    constructive interference of the two signals at the UE;
  \item ``NR'' operates similarly to ``NRP'', but it does not optimize
    the phase shift $\psi_1$;
\item ``HOP'' points the UE beam towards the IRS, thus neglecting the effect of the wall.
\end{itemize}
Significant performance gaps arise when the IRS area is small.  For
$A=1$\,cm$^2$, the ``HOP'' algorithm performs poorly since it points
the UE beam towards the IRS, which provides a very weak signal
compared to that reflected by the wall. Instead, ``NRP'' and ``NR''
perform similarly, since they both steer the UE beam towards the
stronger energy source. However ``NRP'' performs 1--4\,dB better than
``NR'' since the latter does not optimize the phase $\psi_1$.
Instead, for $A=100$\,cm$^2$, ``HOP'' performs identically to the much
more complex ``NRP'' and ``NR''.  We also observe that, for $A=100$
cm$^2$, the curves in Fig.~\ref{fig:wall_on_off}(right) show a 6\,dB
gap w.r.t. those shown in Fig.~\ref{fig:wall_on_off}(left), due to the
gain of the UE ULA with respect to an isotropic antenna.  We conclude that IRSs with area as large as
$100$\,cm$^2$ are required, in order to collect and reflect enough
signal energy to dominate the effect of natural reflectors such as the
plasterboard wall.
On the base of this consideration, in the following we will neglect the contribution of the signal reflected by the plasterboard wall.

\begin{figure}[t]
 \includegraphics[width=0.5\columnwidth]{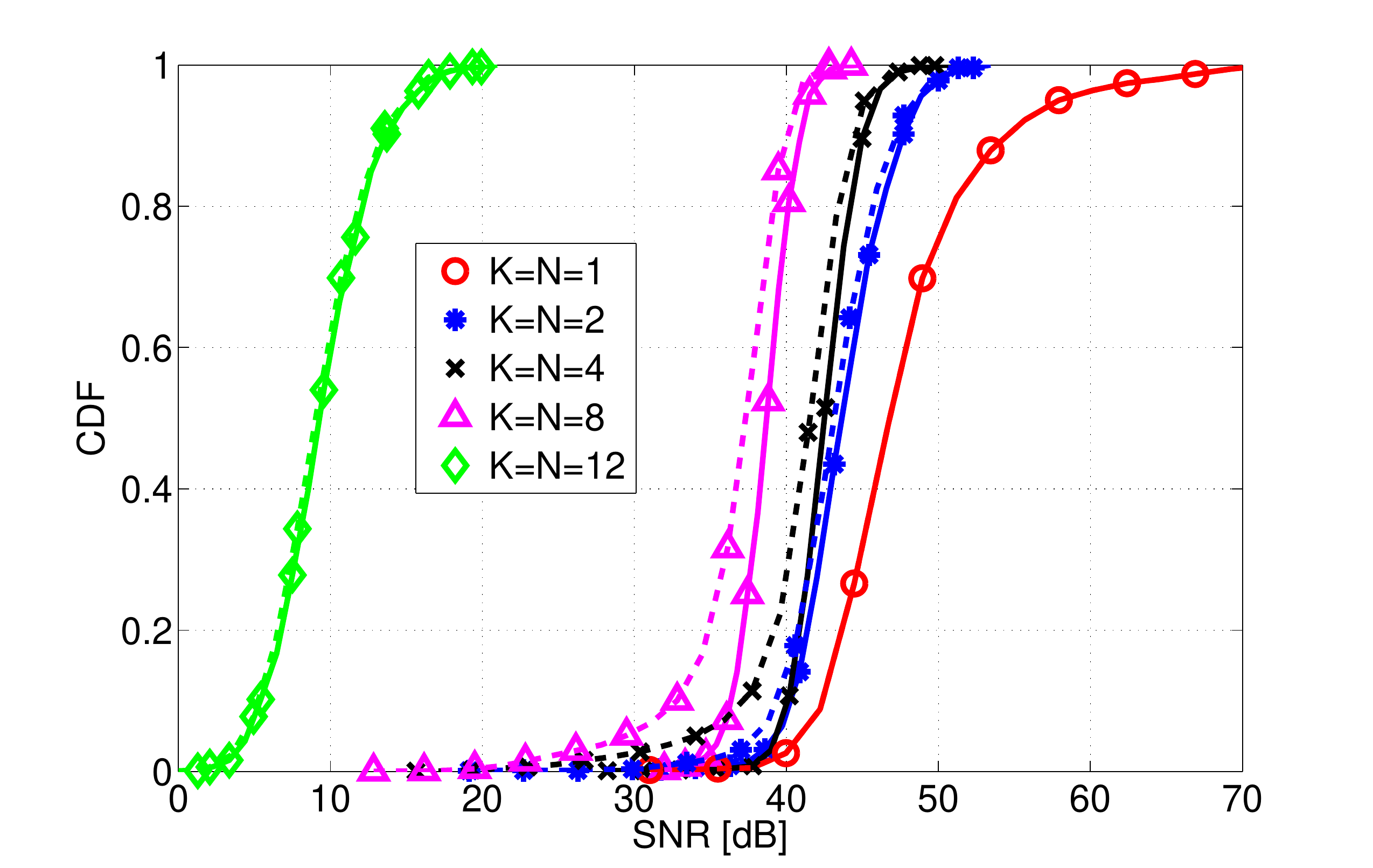}
 \includegraphics[width=0.5\columnwidth]{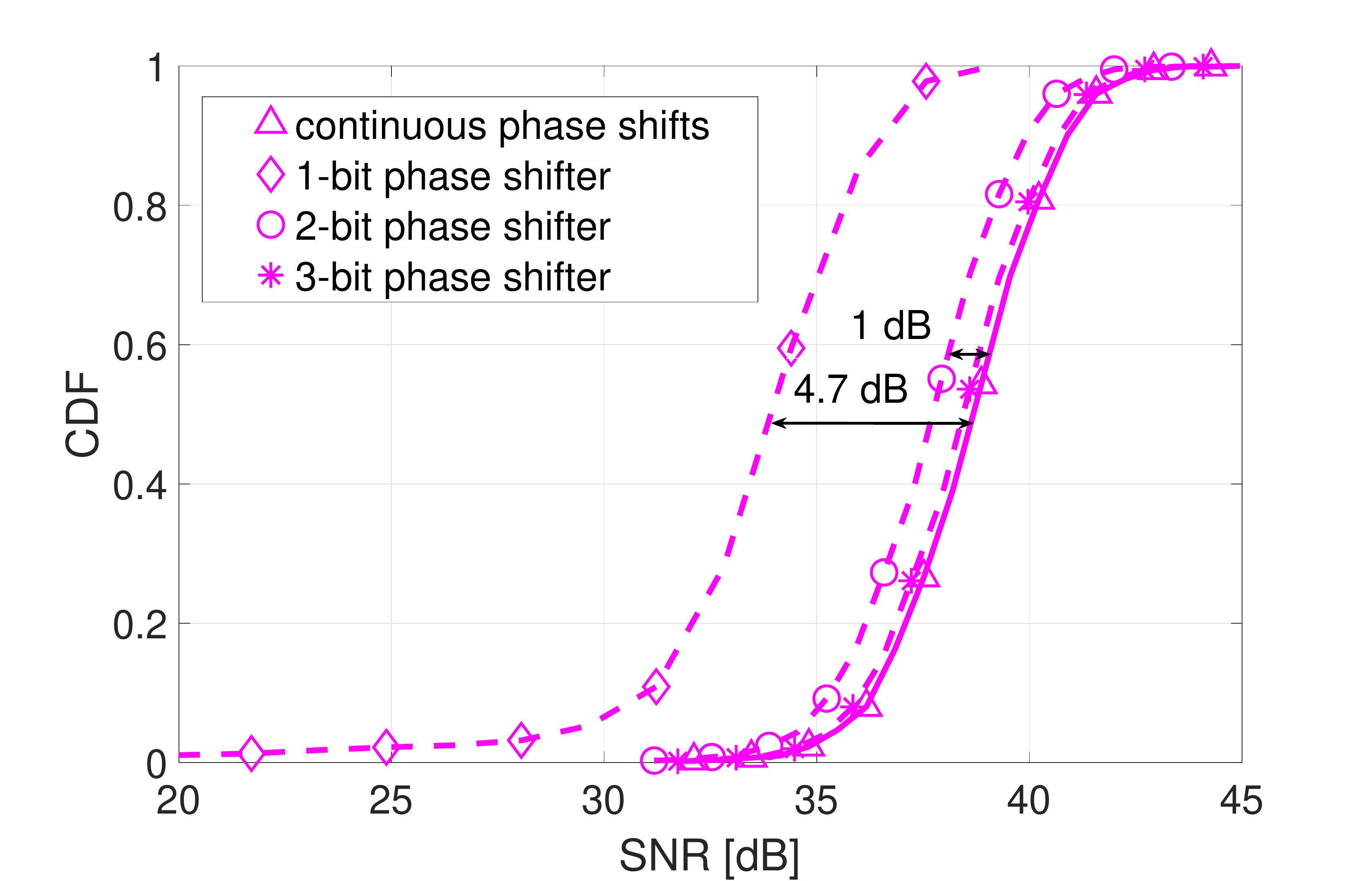}
\vspace{-5ex}
\caption{Cdf of the SNR provided by ``HOP'' for $M_1=32$, $M_2=4$,
   $A=100$\,cm$^2$, in the presence of multipath and shadowing. (Left) $N=K=[1,2,4,8,12]$ assuming continuous phase shifters, (right) $N=K=8$ assuming $b$-bit phase shifters.} \label{fig:Impact}
\end{figure}
\subsection{Impact of the system parameters on the network performance\label{sec:Impact}}
Here we investigate the impact of IRS and UE number on network
performance when shadowing effects and unwanted obstacles are present.
In Fig.~\ref{fig:Impact} (left) we consider a
multi-user scenario and measure the cdf of the SNR for the case $M_1=
32$, $M_2=4$, $A$ = $100$\,cm$^2$, $N=K$, and $K=[1,2,4,8,12]$.  The
results refer to the ``HOP'' algorithm, since ``NR'' and ``NRP'' do
not provide significant performance improvements w.r.t. ``HOP''.

The links from BS to IRSs are assumed to be LoS, whereas the IRS--UE channels
follow the model in~\eqref{eq:chan_2} where $P=2$ NLoS path are
considered. Each NLoS path is characterized by a reflector randomly
positioned in the area $\Uc$ and characterized by a reflection
coefficient $|\rho_{k,n,p}|^2=-10$\,dB.  All the links experience
shadowing effects, i.e., the r.v. $\alpha^{(2)}_{k,n,p}$ are
log-normal distributed with variance $\sigma_{\rm sh}=2$ dB. We also
neglect the reflection due to the plasterboard wall.  For each value
of $K$ two curves are reported. The solid line refers to the case
where the BS has full knowledge of the channel state, including the
shadowing coefficients, the position of the reflectors and of the
users; the dashed line refers to the case where the BS knowledge is
limited to the LoS paths of each IRS-UE channel, i.e., it assumes
$P=0$ and has knowledge of the UEs positions and of the shadowing coefficients $a_{k,n,0}$. In the latter case the BS is unable to
apply the proper ZF filter and, thus to grant an interference free
channel to the UEs.  This clearly entails a performance loss which,
however, is negligible for $K=1,2$, and amounts to about 1\,dB and
2\,dB for $K=4$ and $K=8$, respectively.

Since the total transmitted power $\Pc_t$ is evenly shared among
users, we expect a 3-dB SNR loss as $K$ doubles. In the figure this
can be observed up to a certain value of $K$. However, as $K$ grows,
the SNR loss becomes larger, i.e., it increases to 5\,dB when moving
from $K=4$ to $K=8$, and is as high as 22\,dB when increasing $K$ from 8
to 12. This behavior can be explained as follows. As $K$ and $N$
increase, the distance between adjacent IRSs becomes smaller and so
does the average distance among UEs. When adjacent IRSs are very close
to each other, the BS beam associated to a given UE is not narrow
enough to illuminate a single IRS. Similarly, the
beams reflected by the IRSs are not narrow enough to illuminate a
single UE. In other words, as $K$ increases, many channels become
``almost'' linearly dependent, making the channel matrix ill
conditioned, with many eigenvalues close to 0. Then, the pseudoinverse
$\Hm^+$ in~\eqref{eq:SNR} shows large eigenvalues which have a
detrimental effect on the SNR. In this situation already compromised,
imperfect knowledge of the channel at the BS has negligible impact,
i.e. for $K=12$ the solid and dashed lines are superimposed.

For the same system setting, Fig.~\ref{fig:Impact}
  (right) shows the performance degradation incurred when $b$-bit
  discrete phase shifters are employed. The SNR losses measured when
  $N=K=8$ and $b=1,2,3$ are respectively 4.7\,dB, 1\,dB and 0.2\,dB
  and are consistent with the values reported in~\cite[Table
    I]{Discrete_phase_shift} for an asymptotic (large-$L_n$)
  regime, i.e. 3.9 dB, 0.9 dB and 0.2 dB.
\begin{figure}[t]
 \includegraphics[width=0.5\columnwidth]{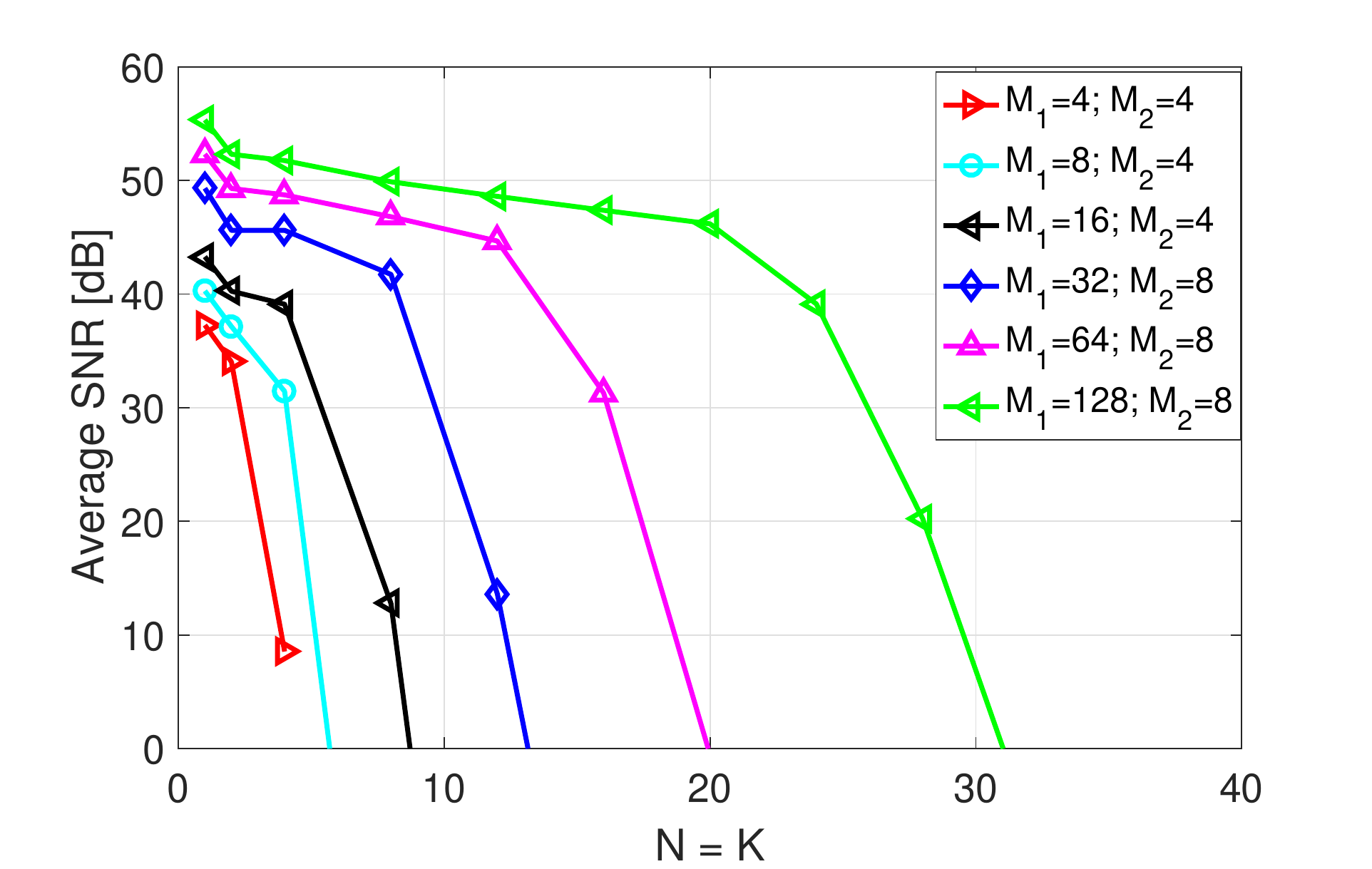}
\includegraphics[width=0.5\columnwidth]{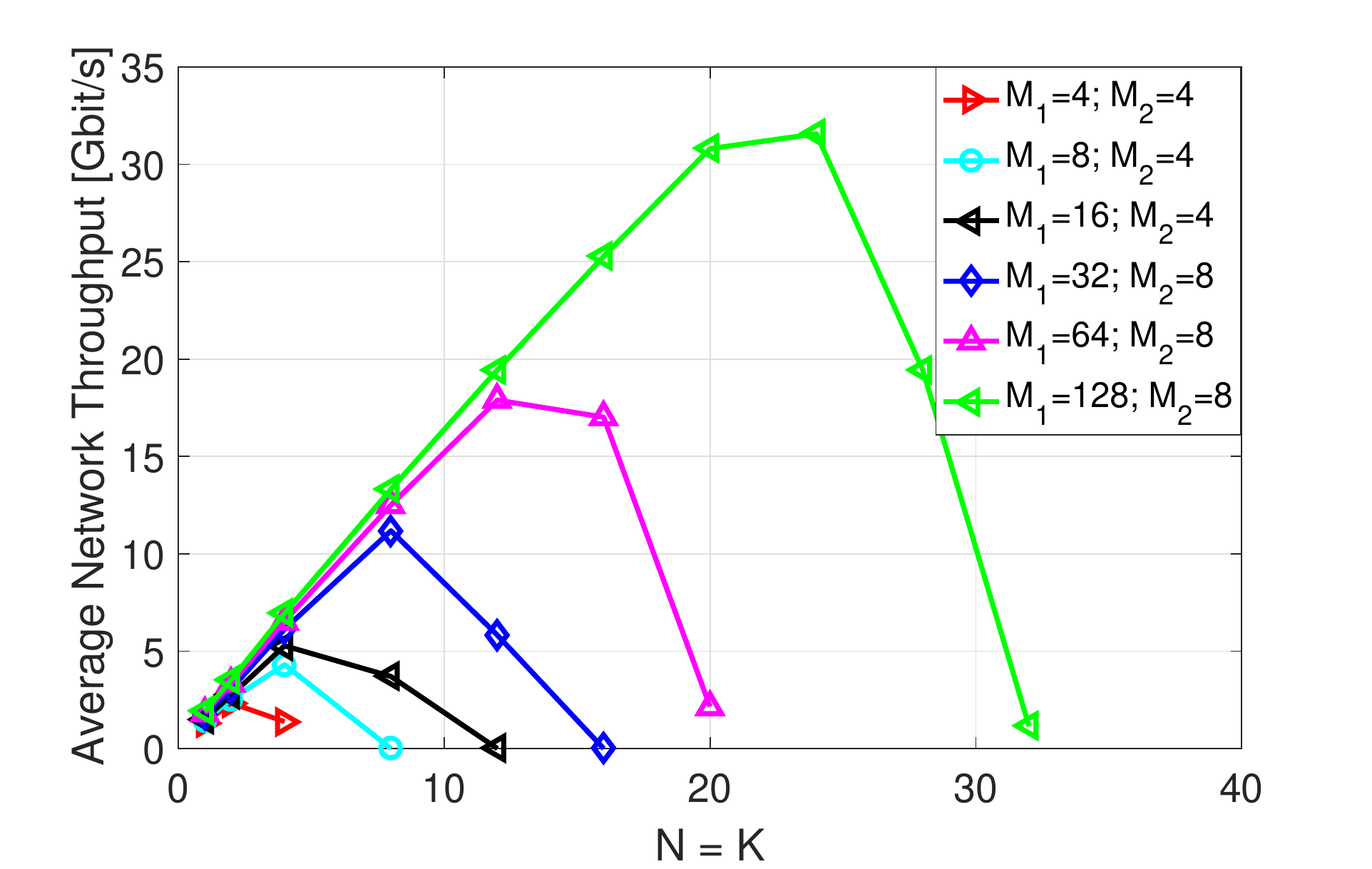}
\vspace{-1cm}
\caption{Average SNR [dB] (Left) and average network throughput R [Gbit/s] (Right) versus the number of
   users, $K=N$, as $M_1$ and $M_2$ vary. The IRSs' area is $A = 100$ cm$^2$.}\label{fig:avgSNR_Throughput}
\end{figure}
The effect of user densification is further investigated in
Fig.~\ref{fig:avgSNR_Throughput}(left) which shows the average SNR
plotted versus $K$, for different values of $M_1$ and $M_2$ in the same setting
of Fig.~\ref{fig:Impact}, where the BS has full knowledge of the channel state.

The curves show an interesting behavior: for small $K$ the SNR slowly
decreases as $K$ increases, with a 3\,dB loss as $K$ doubles; instead,
for large values of $K$, we observe a significant performance
drop. Again, this is explained by observing that for large $K$ the BS
beams are not narrow enough to illuminate a single IRSs, i.e.,
adjacent IRS cannot be ``separated'' by the BS ULA.  Hence the BS-UE
channels cannot be easily orthogonalized.

The overall system performance can also be measured in terms of the
average network throughput, defined as $T = K\cdot B\cdot \EE[R]$,
where $B$ is the signal bandwidth and $R$ is the spectral efficiency
in~\eqref{eq:sp_eff}. Fig.~\ref{fig:avgSNR_Throughput}(right) shows the average
network throughput as a function of the number of supported users, for
different values of $M_1$ and $M_2$. For each pair $(M_1, M_2)$, the
throughput initially increases with $K$. In such a situation the
system load is moderate and the network is able to accommodate more and
more UEs. For $K=K^*$ the system reaches saturation and the throughput
starts falling.  For example, for $M_1=128$ and $M_2=8$, we have $K^*
= 20$ and about 32\,Gbit/s can be achieved. Also, we observe that by
doubling $M_1$ we can double the maximum network throughput.  Indeed,
as $M_1$ increases, the beamwidth of the BS beams decreases and more
IRS can be separated and supported.

As a conclusion, the number of UEs, $K^*$, corresponding to the peak
network throughput can be seen as the maximum order of space division
multiplexing that the IRSs can provide in the particular scenario
investigated. For such number of UEs, in fact, we serve as many users
as possible without compromising the performance, because of the
essentially interference-free channels established from the BS to the
users. For this IRS-assisted environment, the space-division
multiple-access ability of the system is increased by increasing the
size of the arrays at the BS and the UEs, for a sufficiently large
area of the meta-surfaces.  

\section{Conclusions\label{sec:conclusions}}
In this work we tackle the optimization of a SRE composed of a
multiuser wireless network operating in the sub-THz/THz frequency
bands and of a set of IRSs. IRS are employed to improve the BS-UEs
channels when direct BS-UEs LoS links are unavailable.  We considered a channel
model able to capture the main characteristics of sub-THz/THz
propagation such as molecular absorption, multipath, the presence of large solid
objects acting as reflectors, and large-scale fading effects.

Motivated by the extreme sparsity of the sub-THz/THz channel, by the
high gain provided by the transmit and receive antenna arrays, and by
the aim of providing simple solutions for a practical SRE
implementation, we modeled the behavior of each IRS through only two
parameters, namely, the phase-gradient and the phase-shift, abstracting its size and individual components.
According to this model IRSs behave as
electronically steerable reflectors, obeying the generalized Snell
law.

We have shown that such choice, although suboptimal in a general
multiuser scenario, is indeed optimal in many practical relevant cases. Furthermore,
it is extremely appealing since it allows to significantly reduce the
complexity of SRE optimization. Such task is further facilitated by
the adoption, at the BS, of a ZF precoder which orthogonalize UEs
channels and allows to apply a semi-analytic approach to the
optimization algorithms. 
We also provided a set of asymptotic results which provide insight on
the system behavior when the IRSs have large area and the number of
antenna elements at the BS grows large.

Capitalizing on this network model, we have proposed a simple manageable formulation of
the SRE optimization problem, which aim at maximizing the SNR
measured at the UE, whereas the optimization variables are the
electronic rotations and the phase shifts of the IRSs, as well as the
direction of the beams generated by the UE ULAs.
To solve the problem we proposed an algorithm based on Newton-Raphson
method and  a simple heuristic approach based on the Hungarian
algorithm and on a map associating UEs with IRSs.

Our numerical results provide multiple valuable insight.  First, if
surfaces are large enough (e.g. $100$ cm$^2$ in our setup), the
influence of large static reflectors (as walls) can be
neglected. Second, with a sufficiently large number of antennas at the
BS, the heuristic algorithm performs similarly to the more complex
Newton-Raphson approach. Third, as a general rule the number of users
supported by the system depends on the number of antennas at the
BS. Finally, the SNR degradation incurred when discrete
  phase shifters are employed is consistent to that obtained in an asymptotic
  (large-$L_n$) regime. We observe, however, that geometry
also plays an important role since the system performance show a
dramatic drop when the IRS density is so high that they cannot be
angularly separated at the UEs.

In addition to being interesting in themselves, such results further
the high level goal of designing and implementing practical, simple
and efficient IRS aided communication systems working in the THz
frequency bands.  

\appendices
\section{Proof of Proposition \protect\ref{prop:1}}
\label{app:A}
By using the definitions of the matrices $\Hm_n^{(1)}$,
$\bar{\Thetam}_n$, $\Hm_{k,n}^{(2)}$, and $\Hm^{(3)}_n$, the vector
$\fv_k\Herm\widetilde{\Hm}_k$ appearing in~\eqref{eq:yk} can be rewritten as
\begin{eqnarray}
  \fv_k\Herm\widetilde{\Hm}_k &=& \sum_{n=1}^N\sum_{p=0}^P\frac{t_{k,n,p}}{L_n} \left(\onev_{L_n}\Tran\mathord{\otimes} \uv^{(2)}_{k,n,p}\right)\Herm\left(\Id_{L_n}\mathord{\otimes}\Thetam_n\right) (\onev_{L_n}\mathord{\otimes}\uv^{(1)}_{n}){\vv_n^{(1)}}\Herm +\fv_k\Hm_k^{(3)} \non                               
   &=&  \sum_{n=1}^N\sum_{p=0}^Pt_{k,n,p}\left({\uv^{(2)}_{k,n,p}}\Herm \Thetam_n\uv^{(1)}_{n}\right){\vv_n^{(1)}}\Herm+t_k{\vv_k^{(3)}}\Herm\label{eq:Htilde_f}\,,
\end{eqnarray}
where $t_{k,n,p} \triangleq b_{k,n,p}\rho_n c_n^{(1)}c_{k,n,p}^{(2)}$, 
$t_k\triangleq b_k\rho^{\rm wall}c_k^{(3)}$,
$b_{k,n,p}\triangleq \fv_k\Herm\wv_{k,n,p}^{(2)}$, and
$b_k\triangleq \fv_k\Herm\wv_k^{(3)}$.  Furthermore, by recalling the
definitions of $\uv_n^{(1)}$, $\uv^{(2)}_{k,n,p}$ and $\Thetam_n$, and
by assuming uniform illumination of the meta-surface we get
\begin{eqnarray}
  {\uv^{(2)}_{k,n,p}}\Herm \Thetam_n\uv^{(1)}_{n}
  &=& \frac{\ee^{\jj \psi_n}}{L_n}\ee^{\jj \pi \Delta (L_n-1) s_{k,n,p}}\sum_{\ell=1}^{L_n}\ee^{-\jj 2 \pi \Delta (\ell-1) s_{k,n,p}} \non
  &=& \ee^{\jj \psi_n}\frac{\sinc(\Delta L_n s_{k,n,p})}{\sinc(\Delta s_{k,n,p})}\label{eq:f}\,,
\end{eqnarray}
where $s_{k,n,p}= \sin \phi_n^{(1)}-\sin \phi_{k,n,p}^{(2)}-g_n$. Then, 
\begin{equation}
  \fv_k\Herm\widetilde{\Hm}_k
  = \sum_{n=1}^N\sum_{p=0}^P  t_{k,n,p}\ee^{\jj \psi_n}\frac{\sinc(\Delta L_n s_{k,n,p})}{\sinc(\Delta s_{k,n,p})} {\vv_n^{(1)}}\Herm +t_k{\vv_k^{(3)}}\Herm\,.
\end{equation}
  Now, as $L_n$ increases, while the area $A_n$
remains constant, we have $\lim_{L_n\to \infty}\sinc\left(\sqrt{\frac{A_n}{L_n^2\lambda^2}}
  s_{k,n,p}\right)=1$. It follows that
$\lim_{L_n\to \infty}\fv_k\Herm\widetilde{\Hm}_k =\sum_{n=1}^N\sum_{p=0}^P t_{k,n,p}\sinc\left(\sqrt{\frac{A_n}{\lambda^2}} s_{k,n,p}\right)\ee^{\jj \psi_n} {\vv_n^{(1)}}\Herm +t_k{\vv_k^{(3)}}\Herm$.
Since the vector $\fv_k\Herm\widetilde{\Hm}_k$ is the $k$-th row of the matrix $\widetilde{\Hm}$
we can write
\begin{eqnarray}
  \Hm &=&  \lim_{L_1,\ldots,L_N\to \infty}\widetilde{\Hm} = \Mm \Psim{\Vm^{(1)}}\Herm +\Tm{\Vm^{(3)}}\Herm\,,
\end{eqnarray}
$[\Mm]_{k,n}=\rho_nc_n^{(1)}\sum_{p=0}^Pb_{k,n,p}c^{(2)}_{k,n,p}\sinc\left(\sqrt{\frac{A_n}{\lambda^2}} s_{k,n,p}\right)$,
$\Psim = \diag\left(\ee^{\jj \psi_1},\ldots, \ee^{\jj \psi_N}\right)$,
$\Vm^{(1)}=[\vv_1^{(1)},\ldots, \vv_N^{(1)}]$, 
$\Vm^{(3)} = [\vv_1^{(3)}, \ldots,\vv_K^{(3)}]$, and $\Tm = \diag(t_1,\ldots,t_K)$.
Finally, by recalling the expressions for $\fv_k$, $\wv_{k,n,p}^{(2)}$ and $\wv_k^{(3)}$ we obtain
\begin{equation}
 b_{k,n,p} \triangleq \frac{\sinc(\Delta_2M_2(\sin \alpha_k-\sin \zeta_{k,n,p}))}{\sinc(\Delta_2(\sin \alpha_k-\sin \zeta_{k,n,p}))}\,; \quad  b_k \triangleq \frac{\sinc(\Delta_2M_2(\sin \alpha_k-\sin \zeta_k))}{\sinc(\Delta_2(\sin \alpha_k-\sin \zeta_k))}\,.
\end{equation}

\section{Derivation of \protect$\nabla f$ and \protect$\boldsymbol{\Sc}$}\label{app:gradient_and_hessian}
We are interested in computing the gradient and the Hessian of the
term $\| \Hm^{+}\Qm^{1/2} \|_{\rm F}^2$ appearing in~\eqref{eq:SNR}
where $\Hm=\Mm \Psim{\Vm^{(1)}}\Herm+\Tm{\Vm^{(3)}}\Herm$.  First of
all, we define $\Km = \Hm \Hm\Herm$ and we observe that
$\| \Hm^{+}\Qm^{1/2} \|_{\rm F}^2 = \trace\{ \Km^{-1}\Qm\}$.  Let
$\deltav=[\delta_1,\ldots,\delta_N]\Tran$,
$\psiv=[\psi_1,\dots,\psi_N]\Tran$, and
$\alphav=[\alpha_1,\dots,\alpha_K]\Tran$, be the vectors of
variables to be optimized. Then we can define
$f(\deltav,\psiv, \alphav) \triangleq \trace\{ \Km^{-1}\Qm \}$.  Let
$x$ be a generic argument of the function $f(\cdot)$, and let $\Am$ be
a matrix. Then we define
$\frac{\partial \Am}{\partial x} = \Am_{(x)}$. Also, the first
derivative of $f(\cdot)$ w.r.t. $x$ is given by
\begin{eqnarray}
    \frac{\partial f}{\partial x}
    &=&\sum_{i,j}\frac{\partial\trace\{ \Km^{-1}\Qm\} }{\partial K_{i,j}}\frac{\partial K_{i,j}}{\partial x}
    =\sum_{i,j}\trace\left\{\frac{\partial\trace\{ \Ym^{-1}\Qm \} }{\partial\Ym}\Big|_{\Ym=\Km} \frac{\partial \Km}{\partial K_{i,j}}\right\}\frac{\partial K_{i,j}}{\partial x}\non
    &=&-\sum_{i,j}\trace\left\{ \left(\Km^{-1}\Qm\Km^{-1} \right)\Tran \Jm^{(i,j)}\right\}\frac{\partial K_{i,j}}{\partial x}\label{eq:first_derivative}\,,
  \end{eqnarray}
  where the last equality comes from
  \cite[Eq. (121)]{Cookbook}. In~\eqref{eq:first_derivative}, $\Ym$ is
  a matrix whose entries are independent variables, and
  $\Jm^{(i,j)}=\frac{\partial \Km}{\partial K_{i,j}}$ represents the
  structure of the matrix $\Km$.  The matrix $\Km$ is complex
  Hermitian, thus $[\Jm^{(i,j)}]_{m,n} = 0$ for $(m,n)\neq (i,j)$ and
  $(m,n)\neq (j,i)$. Clearly $[\Jm^{(i,j)}]_{i,j} = 1$,
  whereas\footnote{Here, in order to handle complex differentiation of
    non analytic functions we use the definition of Wirtinger
    derivatives~\cite{Remmert}.}
  $[\Jm^{(i,j)}]_{j,i} = \frac{\partial K_{j,i}}{\partial K_{i,j}} =
  \frac{\partial K_{i,j}^*}{\partial K_{i,j}} = 0$. Then, from~\eqref{eq:first_derivative} we obtain
\[ \frac{\partial f}{\partial x}
    =-\sum_{i,j}[\Km^{-1}\Qm\Km^{-1}]_{j,i}\frac{\partial K_{i,j}}{\partial x}
    =-\trace\left\{ \Km^{-1}\Qm\Km^{-1}\frac{\partial \Km}{\partial x}\right\}\non
    =-\trace\left\{ \Zm \Km_{(x)}\right\}\,,
\]
  where $\Zm \triangleq \Km^{-1}\Qm\Km^{-1}$.
    Now let $y$ be another argument of the function $f(\deltav,\psiv,\varphiv)$.
  The second mixed derivative of $f(\cdot)$ is given by
  \begin{eqnarray}
    \frac{\partial^2 f}{\partial x \partial y}
    &=&-\frac{\partial}{\partial y}\trace\left\{ \Zm\Km_{(x)}\right\}
    =-\trace\left\{ \frac{\partial\Zm}{\partial y}\Km_{(x)}+\Zm\frac{\partial}{\partial y}\Km_{(x)} \right\}\,.
  \end{eqnarray}
  Now observe that $\frac{\partial\Zm}{\partial y} = -2 \Zm\Km_{(y)}\Km^{-1}$. Thus we obtain
  \begin{eqnarray}
    \frac{\partial^2 f}{\partial x \partial y} 
    &=&-\trace\left\{ -2 \Zm\Km_{(y)}\Km^{-1}\Km_{(x)}+\Zm\Km_{(xy)} \right\}
    =\trace\left\{ \Zm (2 \Km_{(y)}\Km^{-1}\Km_{(x)}-\Km_{(xy)}) \right\}
  \end{eqnarray}
  where $\Km_{(xy)}=\frac{\partial}{\partial y}\Km_{(x)}$. The Hessian of $f(\cdot)$ is then defined in terms  of the derivatives of $\Km$ 
  \begin{equation} \Km_{(x)} =\frac{\partial}{\partial x}\Hm\Hm\Herm = \Hm_{(x)}\Hm\Herm + \Hm\Hm_{(x)}\Herm \label{eq:partial_x}\,,\end{equation}
  and, thus
   $ \Km_{(xy)} =\Hm_{(xy)}\Hm\Herm +\Hm_{(x)}\Hm_{(y)}\Herm + \Hm\Hm_{(xy)}\Herm +\Hm_{(y)}\Hm_{(x)}\Herm \label{eq:partial_xy}$.
  The derivatives of $\Hm$ are easy to obtain
  from~\eqref{eq:H_mat}. In particular the matrix $\Mm$ depends on
  both $\deltav$ and on $\alphav$, the matrix $\Psim$ depends on
  $\psiv$ only, and $\Tm$ depends on $\alphav$ only. The obtained
  expressions are quite cumbersome and, for simplicity, are not reported here. 

\section{Proof of Proposition \protect\ref{prop:2}\label{app:C}}
We start from \eqref{eq:Htilde_f} and we specialize it to the case $K = N = 2$,
$P = 0$ and no wall reflection (i.e., $t_k=0$ for all $k$).  We can write the
overall channel matrix from the BS to the 2 UEs as $\widetilde{\Hm} =
\widetilde{\Mm} {\Vm^{(1)}}\Herm$, with $\Vm^{(1)}$ defined as in
Prop. \ref{prop:1}, and $\widetilde{\Mm}$ a $2 \times 2$ matrix, with
$(k,n)$ element $\widetilde{m}_{k,n} = t_{k,n} {\uv^{(2)}_{k,n}}\Herm \Thetam_n\uv^{(1)}_{n} = t_{k,n} {\uv^{(2)}_{k,n}}\Herm \widetilde{\uv}_n$ 
where we have dropped the subscript $p$ and we have defined the
length-$L_n$ norm-1 vector $\widetilde{\uv}_n$, which satisfies the
equimodular property $|(\widetilde{\uv}_n)_i| = 1/\sqrt{L_n}$.
We define the optimal value of the IRS phase shifts as the one that
minimizes $ \|\widetilde{\Hm}^{+}\Qm^{1/2} \|_{\rm F}^2$, where
$\widetilde{\Hm}^{+} = \widetilde{\Hm}\Herm
(\widetilde{\Hm}\widetilde{\Hm}\Herm)^{-1} $ is the pseudo-inverse of
$\widetilde{\Hm}$. In Proposition \ref{prop:2}, we suppose for simplicity $\Qm = \Id_2$,
although the generalization is straightforward. When $M_1 \rightarrow
\infty$, as in Sect. \ref{sec:BS_ant_asymp}, $\Vm^{(1)}$ tends to a unitary
matrix, so that, similarly to \eqref{eq:limitMinf}, $
\|\widetilde{\Hm}^{+} \|_{\rm F}^2 \stackrel{M_1\to
  \infty}{\longrightarrow}
\trace\left\{\left(\widetilde{\Mm}\widetilde{\Mm}\Herm\right)^{-1}\right\}\,.
$

 For $L_n \rightarrow \infty$, the spatial signatures
 $\{\uv^{(2)}_{k,n}\}_{k,n=1}^2$ become orthogonal, provided that all
 users are angularly separated. Thus, for IRS $n$, the space of useful
 signal is the bidimensional space spanned by the orthonormal basis
 $\{\uv^{(2)}_{1,n},\uv^{(2)}_{2,n}\}$. So, we can write
 $\widetilde{\uv}_n$ as
\begin{equation}
\widetilde{\uv}_n = \cos \theta_n \cos \phi_n \uv^{(2)}_{1,n} + \cos \theta_n \sin \phi_n \uv^{(2)}_{2,n} +  \widetilde{\uv}_n^{\perp}
\end{equation}
where $\widetilde{\uv}_n^{\perp}$ is the component of
$\widetilde{\uv}_n$ orthogonal to the useful signal space. Defining
$\gamma_n = \cos \theta_n$, $\kappa_n = \cos \phi_n$,
$\sigma_n = \sin \phi_n$ and $\mu_{k,n} = |t_{k,n}|$, we can
reformulate the optimization problem as the maximization of function
$f$ given by
\begin{equation}
f = \trace\left\{\left(\widetilde{\Mm}\widetilde{\Mm}\Herm\right)^{-1}\right\}^{-1} = \frac{(\mu_{11} \mu_{22} \kappa_1 \sigma_2 - \mu_{12} \mu_{21} \sigma_1 \kappa_2 )^2}{1/\gamma_1^2 (\mu_{11}^2 \kappa_1^2 + \mu_{21}^2 \sigma_1^2) + 1/\gamma_2^2 (\mu_{12}^2 \kappa_2^2 + \mu_{22}^2 \sigma_2^2)} \,.
\end{equation}
Now, we solve the optimization problem without considering the equimodular condition on $\widetilde{\uv}_n$. First, the maximum of $f$ is obtained for $\gamma_1 = \gamma_2 = 1$, i.e.,  $\widetilde{\uv}_n$ belongs to the useful signal space, a pretty obvious fact. To maximize $f$ with respect to $\phi_n$, $n=1,2$, we set the gradient $\nabla f$ to zero. Writing $f = f_1/f_2$, we have for $n = 1,2$
\[
\frac{\partial f}{\partial \phi_n} = \frac{\partial f_1}{\partial \phi_n} \frac1{f_2} - \frac{\partial f_2}{\partial \phi_n} \frac{f_1}{f_2^2} = 0 \,\,\,\Longrightarrow\,\,\, \frac{\partial f_1}{\partial \phi_n} =  \frac{\partial f_2}{\partial \phi_n} \frac{f_1}{f_2}.
\]
We obtain the following two equations (for $f_1 > 0$, since $f_1 = 0$ gives a minimum of $f$):
\begin{eqnarray*}
\mu_{12} \mu_{21} \sigma_1 \sigma_2 (\mu_{22}^2 + \zeta_1) + 
\mu_{11} \mu_{22} \kappa_1 \kappa_2 (\mu_{12}^2 + \zeta_1) &=& 0\\
\mu_{11} \mu_{22} \sigma_1 \sigma_2 (\mu_{21}^2 + \zeta_2) + 
\mu_{12} \mu_{21} \kappa_1 \kappa_2 (\mu_{11}^2 + \zeta_2) &=& 0\,.
\end{eqnarray*} 
having defined
$\zeta_1 = \mu_{11}^2 \kappa_1^2 + \mu_{21}^2 \sigma_1^2$ and
$\zeta_2 = \mu_{12}^2 \kappa_2^2 + \mu_{22}^2 \sigma_2^2$. The above
equations are satisfied if $\sigma_1 \sigma_2=0$ and
$\kappa_1 \kappa_2=0$. This yields two points in the first quadrant,
i.e. $(\phi_1, \phi_2) = (0, \pi/2)$ and
$(\phi_1, \phi_2) = (\pi/2, 0)$. The first point corresponds to
assigning user 1 to IRS 1 and user 2 to IRS 2, while the second
assigns user 2 to IRS 1 and user 1 to IRS 2. Instead, if
$\sigma_1 \sigma_2 \neq 0$ and $\kappa_1 \kappa_2 \neq 0$, we can
solve both equations above for
$\frac{\sigma_1 \sigma_2}{\kappa_1 \kappa_2}$ and equate the
solutions. By doing this, after a little bookkeeping, we obtain the
following equation:
\[
\mu_{11}^2 \mu_{21}^2 (\mu_{22}^2 - \mu_{12}^2) \zeta_1 + \mu_{12}^2 \mu_{22}^2 (\mu_{11}^2 - \mu_{21}^2) \zeta_2 + (\mu_{11}^2 \mu_{22}^2 - \mu_{21}^2 \mu_{12}^2) \zeta_1 \zeta_2 = 0
\]
But, if $\mu_{11} > \mu_{21}$ and $\mu_{22} > \mu_{12}$, all
coefficients of $\zeta_1$ and $\zeta_2$ are positive and, since
$\zeta_n >0$, $n = 1,2$, the above equation does not have any
solution. Analogously if $\mu_{11} < \mu_{21}$ and
$\mu_{22}< \mu_{12}$. Thus, in such conditions, the only two
stationary points are those corresponding to IRS-user assignments, and
one of the two must be the global maximum.  The global maximum is the
first point if
\[
\frac1{\mu_{11}^2} + \frac1{\mu_{22}^2} < \frac1{\mu_{12}^2} + \frac1{\mu_{21}^2}
\] 
otherwise the global maximum is the second point. The obtained optimal
IRS phase shifts are equal to the solution of the optimization problem
in \eqref{eq:optimize_simple2}.  

\bibliographystyle{IEEEtran}
\bibliography{refs}

\end{document}